%% file: manuscript.tex
\documentclass[10pt,aps,prd,reprint,nofootinbib,floats,floatfix,amsfonts,amssymb,amsmath,preprintnumbers,notitlepage,superscriptaddress,longbibliography]{revtex4-1}
\pdfoutput=1
\usepackage{aas_macros}
\usepackage{subfigure}
\usepackage{amsfonts}
\usepackage{amsmath}
\usepackage[utf8]{inputenc}
\usepackage[british]{babel}
\usepackage[Symbolsmallscale]{upgreek}
\usepackage[protrusion=true,tracking=true,kerning=true,spacing=true,final,babel=true]{microtype}
\usepackage{physics}
\usepackage{xcolor}
\usepackage{graphicx}
\usepackage[final]{hyperref}
\usepackage{nth}

\def\be{\begin{equation}}
\def\ee{\end{equation}}

\begin{document}
\title{Realistic sensitivity curves for pulsar timing arrays}

\author{Jeffrey~S.~Hazboun}
\affiliation{University of Washington Bothell, 18115 Campus Way NE, Bothell, WA 98011, USA}

\author{Joseph~D.~Romano}
\affiliation{Department of Physics and Astronomy, Texas Tech University, 
Lubbock, TX 79409-1051, USA}

\author{Tristan~L.~Smith}
\affiliation{Department of Physics and Astronomy, Swarthmore College,
500 College Ave., Swarthmore, PA 19081, USA}

\date{\today}

\input{./manuscript/abstract}
\maketitle

\input{./manuscript/introduction}
\input{./manuscript/timing}
\input{./manuscript/response}
\input{./manuscript/sensitivity}
\input{./manuscript/discussion}
\input{./manuscript/acknowledgements}

\appendix
\input{./manuscript/appendix}

\bibliography{pta_sensitivity}

\end{document}

%% file: manuscript/abstract.tex
\begin{abstract}
We construct realistic sensitivity curves for pulsar timing array
searches for gravitational waves, incorporating both red and white
noise contributions to individual pulsar noise spectra, as well as the
effect of fitting to a pulsar timing model.  We demonstrate the method
on both simulated pulsars and a realistic array consisting of a subset
of NANOGrav pulsars used in recent analyses.  A comparison between the
results presented here and exact sensitivity curves shows agreement to
tens of percent.  The resulting sensitivity curves can be used to
assess the detectability of predicted gravitational-wave signals in
the nanohertz frequency band in a fraction of the time that it would 
take to compute the exact sensitivity curves.  
\end{abstract}

%% file: manuscript/introduction.tex
\section{Motivation}
\label{s:introduction}

Pulsar timing arrays (PTAs) are poised to make the first detection of
nanohertz gravitational waves (GWs) in the next 2-5 years
\cite{2013CQGra..30v4015S,rsg15,2016ApJ...819L...6T,2017MNRAS.471.4508K}. 
These galactic-scale GW detectors use the
correlated times of arrival (TOAs) from millisecond pulsars to search
for GWs \cite{saz78, det79, fb90}.  The recent inception of
observational relativity by the advanced LIGO and VIRGO ground-based
detectors~\cite{Abbott:2016blz, LIGOScientific:2018mvr}
and the multi-messenger observations of binary
neutron stars~\cite{TheLIGOScientific:2017qsa} 
have drastically changed our understanding of
stellar-mass compact objects. PTAs are poised to complement these
observations by observing GWs from binary systems comprised of
super-massive black holes (SMBHs) in the centers of distant galaxies. 

\begin{figure}[h!tbp]
\centering
\includegraphics[width=\columnwidth]{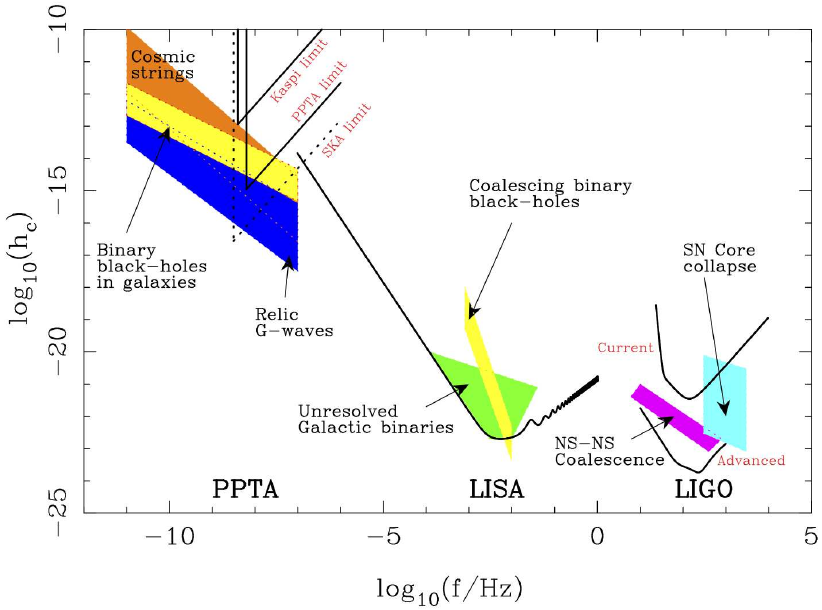}
\caption{Sensitivity curves for different GW observations and 
the predicted spectra of various GW sources.
Note, in particular, the (over) simplicity of the PTA sensitivity 
curves relative to those for LISA and LIGO.
The goal of our paper is to construct more realistic PTA sensitivity curves.
(Figure taken from \cite{2011ASSP...21..229H}.)}
\label{f:pta_wedge}
\end{figure}

A common tool used to assess the observability of GW sources across
the spectrum are detection sensitivity curves 
(see, e.g., \cite{Moore:2014lga, Cornish:2018dyw} and Figure~\ref{f:pta_wedge}).
These curves are basic ``figures of merit," 
constructed by the developers of GW
observatories to assess the sensitivity of current detectors and to
predict the sensitivity of future, next-generation detectors. The
wider astrophysics community uses detection sensitivity curves as an
initial estimate of the ability of a given detector to observe GWs
from a particular source. 

While detailed sensitivity curves for extant detectors are usually
published for each observation run, those for PTAs are often
simplified \cite{2011ASSP...21..229H, Thrane:2013oya},
only including identical white-noise components and often
assuming that all pulsar observation epochs are evenly spaced and
have the same baseline of observations.  
When drawn, these curves are often cut-off at the timespan of the 
observations and do
not include important {\em insensitivities} at frequencies of $1/{\rm
yr}$ and $2/{\rm yr}$, due to fitting for a pulsar's astrometric
parameters (Figure~\ref{f:pta_wedge}).

It has long been known that the fit to a pulsar's timing model acts as
a filter function \cite{1980ApJ...237..216C, Blandford:1984a},
absorbing frequencies in the pulsar timing data in a predictable
manner.  These effects have been studied in the context of searches
for GWs \cite{Cutler:2013aja, 2016ApJ...819L...6T,
Blandford:1984a}. 
Reference \cite{Madison:2012as} go one step further, showing
how one can mitigate for losses in sensitivity using very-long-baseline 
interferometry to localize pulsars sky locations without
explicitly fitting for their positions using the timing data. 

Modern PTA data analysis strategies and algorithms are designed with
this complication of the timing model fit in mind
\cite{2007PhDT........14D, vanHaasteren:2008yh, 2011MNRAS.418..561C,
vanHaasteren:2012hj, Chamberlin:2014ria, 2012MNRAS.423.2642L, 2016ASPC..502...19L}.
This formalism was used e.g., in
\cite{Moore:2014eua} to study PTA sensitivity curves for
deterministic and stochastic sources of GWs, calculating 
sensitivity curves both analytically and numerically,
using frequentist and Bayesian methods.
The approach in \cite{Moore:2014eua} 
is similar in spirit to ours in that they
start from the same likelihood function as we do
(Section~\ref{s:fitting}), and they use properties of the 
expected signal-to-noise ratios for deterministic and
stochastic GW signals to start to incorporate the effect of
timing model fits.
Our analysis differs from theirs in that we explicitly 
identify a component of the likelihood function 
that encodes both the noise power spectral density in a
given pulsar's data set and the effects of the timing model fit.  
This information is combined with known sources of realistic noise in
pulsar timing data, including time-correlated (red) noise, to
construct sensitivity curves for individual pulsars.  
(Reference~\cite{2016MNRAS.457.4421C} also discusses 
the effect of red noise on the sensitivity of pulsar timing 
searches for GWs, using a Fisher matrix calculation to 
estimate the errors.)
For an array of pulsars, we use the expected signal-to-noise ratio 
of detection statistics for both deterministic and stochastic 
GW signals to construct effective sensitivity curves for the 
whole array.

\subsection{Plan of paper}
\label{s:outline}

In Section~\ref{s:timing}, we describe the basic formalism 
underlying pulsar timing analyses---i.e., timing 
residuals, timing models, and the effect of fitting to a 
timing model.
This leads us to timing-model-marginalized residuals
and their associated transmission functions, which play
a key role in the subsequent construction of detection
sensitivity curves.
In Section~\ref{s:response}, we describe in detail the 
response of pulsar timing measurements to both
deterministic and stochastic GWs.
Then, in Section~\ref{s:sensitivity}, we introduce detection 
statistics for both types of signals. 
The expressions for their corresponding expected 
signal-to-noise ratios allow us to read off an 
effective strain-noise power spectral density for the PTA, 
which has the interpretation of a detection sensitivity curve.
As an application of our analysis, we construct sensitivity 
curves for the NANOGrav 11-yr pulsars using realistic 
noise properties and
timing model fits, and compare our predicted sensitivities
to published upper limits.
We conclude in Section~\ref{s:discussion}.  
We also include Appendix~\ref{s:appendix},
in which we cast the results of an early seminal paper
\cite{Blandford:1984a} into the more modern notation used in recent
pulsar timing analyses.

The calculations provided in this work are packaged in a Python package 
available on the Python Package Inventory (PyPI) and GitHub. 

%% file: manuscript/timing.tex
\section{Pulsar timing analyses}
\label{s:timing}

Here we review the formalism underlying pulsar timing analyses used in
GW searches.  Readers interested in more details should see
\cite{Blandford:1984a, 2007PhDT........14D, Ellis:2012zv, vanHaasteren:2012hj,  Chamberlin:2014ria}.

\subsection{Times of arrival and timing residuals}
\label{s:TOAs_residuals}

Let us start with a single pulsar.
The measured pulse times of arrival (TOAs) consist of three parts:%
\footnote{To simplify the notation, we have not included indices
to label the particular pulsar ($I=1,2,\cdots,N_{\rm p}$), the 
individual TOAs ($i=1,2,\cdots,N$), or the timing model 
parameters ($a=1,2,\cdots,N_{\rm par}$).
If one wants to include those indices explicitly, one should
write $t_{Ii} = t_{Ii}^{\rm det}(\xi_a) + n_{Ii} + h_{Ii}$.}
\be
t = t^{\rm det}(\xi) + n + h\,.
\label{e:TOAs}
\ee
The first term gives the expected TOAs due to deterministic
processes, which depend on intrinsic properties of the 
pulsar (e.g., its spin period, period derivative, ...),
extrinsic properties of the pulsar (e.g., its sky location, 
proper motion, distance from the solar system barycenter, ...),
and processes affecting the pulse propagation (e.g., 
disperion delays due to the interstellar medium, relativistic 
corrections, ...).
The timing model parameters are denoted by $\xi$.
The second term is (stochastic) noise intrinsic to the pulsar or 
to the measurement process itself.
The third term is a perturbation to the pulse arrival times 
induced by GWs, 
which in general will have contributions from both deterministic 
and stochastic sources, $h=h^{\rm det} + h^{\rm stoch}$.

Timing residuals are then defined by subtracting the 
expected TOAs (predicted by the timing model for an 
initial estimate of the model parameters $\xi_0$)
from the measured TOAs:
\be
\delta t \equiv t-t^{\rm det}(\xi_0)
= M\,\delta\xi + n+h\,,
\label{e:residuals}
\ee
where
\be
M\equiv \left(\frac{\partial t^{\rm det}}{\partial\xi}\right)\bigg|_{\xi=\xi_0}
\label{e:M}
\ee
is the {\em design matrix}.
The above expression for $\delta t$ is obtained by Taylor expanding 
the timing model $t^{\rm det}(\xi)$ around the initial parameter 
estimates $\xi_0$, assuming that the initial 
estimates are close enough to the true values that only 1st-order 
terms in the parameter deviations $\delta\xi$ are needed in the expansion.
The design matrix $M$ is a rectangular matrix of 
dimension $N\times N_{\rm par}$, with components $M_{ia}$.
Each column of the design matrix encodes the linearized fit to one 
parameter in the timing model. 

\subsection{Fitting to a timing model}
\label{s:fitting}

From the form of \eqref{e:residuals}, one sees that errors
$\delta\xi$ 
in our orignal estimate $\xi_0$ of the timing model parameters 
lead to deterministic features in the timing residuals.
For example, an error in the pulse period leads to timing residuals that 
grow linearly with time, $\delta t\sim t$, while an error 
in the period derivative leads to residuals that grow 
quadratically with time, $\delta t\sim t^2$.
Thus, we can improve our estimates of the timing model
parameters by fitting for $\delta\xi$ in our linear timing 
model for the residuals.

This can be done in two ways, both of which take the likelihood
function 
\be
\begin{aligned}
&p(\delta t|\delta\xi, C_n, C_h, \theta) \propto
\\ 
&\exp\left[-\frac{1}{2}(\delta t - M \delta\xi - h(\theta))^T C^{-1}
(\delta t - M \delta\xi - h(\theta))\right]
\label{e:likelihood}
\end{aligned}
\ee
as the starting point.
In the above expression, 
\be
C\equiv C_n + C_h
\ee
is the noise covariance matrix, 
which has contributions from both detector noise $C_n$ 
(i.e., noise intrinsic to the pulsar and from the measurement process) 
and a potential GW background $C_h$.
The term $h(\theta)$
are the timing residuals induced by a deterministic GW source
(e.g., the expected waveform from an individual SMBH binary
parametrized by $\theta$).

(i) The first approach to fitting to the timing model 
is to {\em maximize} the likelihood function
with respect to the parameter deviations $\delta\xi$.
Since $\delta\xi$ appears linearly in the expression for
the timing residuals (quadratically in the argument
of the exponential), the maximization is easy to do.
One obtains the standard result
\be
\delta\xi_{\rm ML} = (M^TC^{-1}M)^{-1}M^TC^{-1}\delta t\,.
\ee
From these maximum-likelihood estimates, we can then form
{\em post-fit} residuals
\begin{align}
&\delta t^{\rm post} 
\equiv\delta t - M\,\delta\xi_{\rm ML} = R\,\delta t\,,
\\
&R\equiv 1 - M(M^T C^{-1} M)^{-1}M^T C^{-1}\,.
\end{align}
Note that $R$ is an $N\times N$ matrix
that implements the fit to the linear timing model; it 
depends in general on both the timing model (via $M$) 
and the detector noise (via $C$).
One can show that $R$ is a projection operator ($R^2=R$), 
and hence not invertible.

(ii) The second approach to fitting to the timing model is 
to {\em marginalize} the 
likelihood function over the parameter deviations $\delta\xi$, 
assuming flat priors for $\delta\xi$.
The result of this marginalization is the timing-model-marginalized
(TMM) likelihood function~\cite{vanHaasteren:2008yh,vanHaasteren:2012hj}
\be
\begin{aligned}
&p(\delta t|C_n, C_h, \theta)\propto
\\
&\exp\left[
-\frac{1}{2}(\delta t - h(\theta))^TG(G^TCG)^{-1}G^T
(\delta t-h(\theta))\right]\,,
\label{e:likelihood_TMM}
\end{aligned}
\ee
where $G$ is an $N\times(N-N_{\rm par})$ matrix constructed
from a singular-value decomposition of the design matrix
\be
M= U S V^T\,, \quad U = (F,G)\,.
\label{e:M_svd}
\ee
Note that $G$ depends only on the timing model (via $M$) 
and not on the noise.
In terms of components, $G\equiv G_{i\alpha}$, where 
$\alpha=1,2,\cdots,N-N_{\rm par}$.
Using $G$, one can construct associated TMM residuals 
\be
r \equiv G^T\,\delta t\,,
\label{e:r}
\ee
which are orthogonal to the timing model.
Since $U$ is a unitary matix, it follows that 
$[G^T G]_{\alpha\beta}=\delta_{\alpha\beta}$.
For white noise (i.e., $C$ proportional to the 
identity matrix), we have the identitiy $R=GG^T$.

Although both approaches for fitting to the timing model
have been used in the past, in this paper we will use the
second approach, given that it is the one used most often
for current pulsar timing array searches for GWs.

\subsection{Transmission functions}
\label{s:transmission}

\begin{figure*}[h!tbp]
\centering
\subfigure[]
{\includegraphics[width=0.48\textwidth]{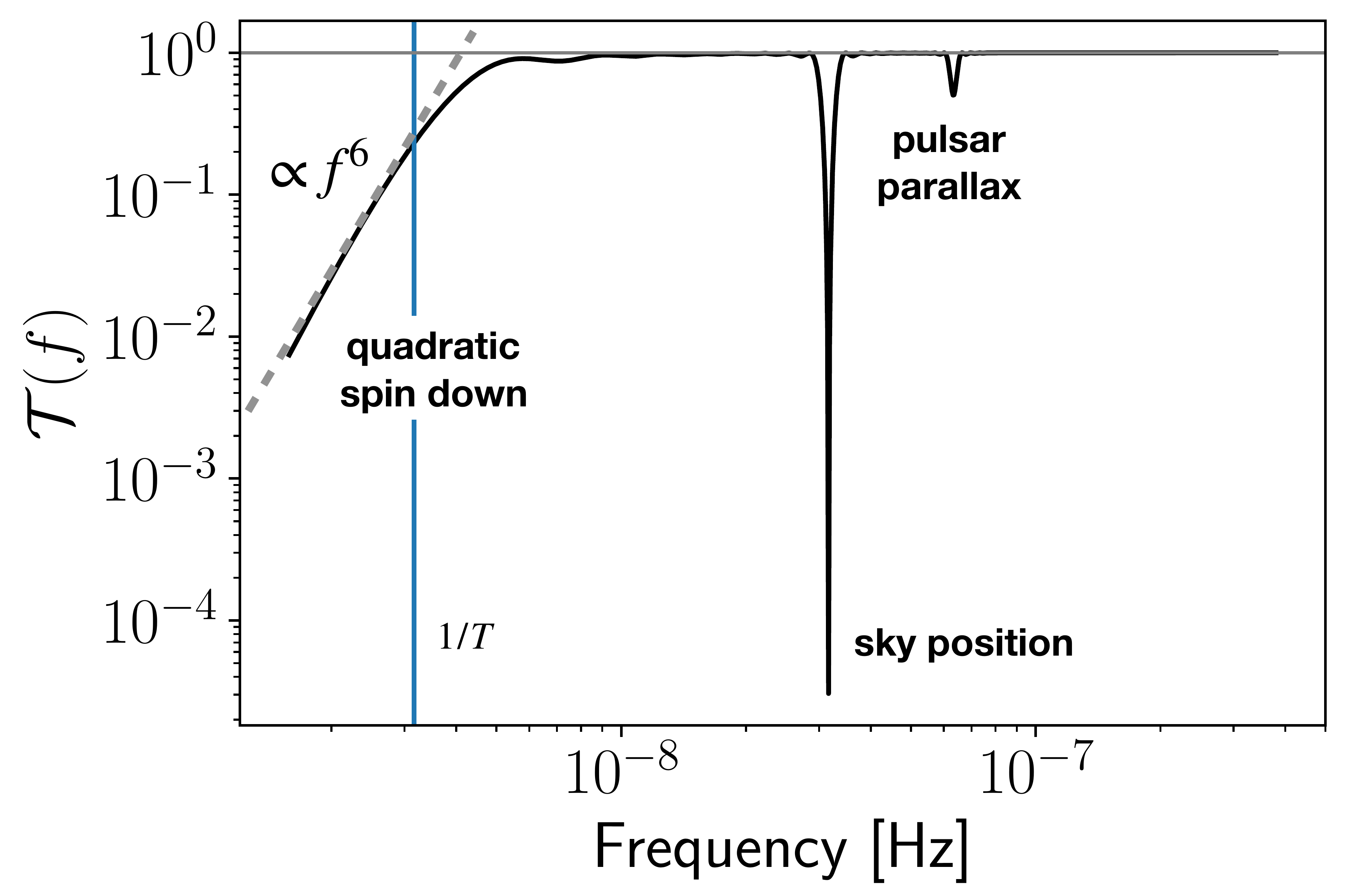}
\label{f:transmission}}
\subfigure[]
{\includegraphics[width=0.48\textwidth]{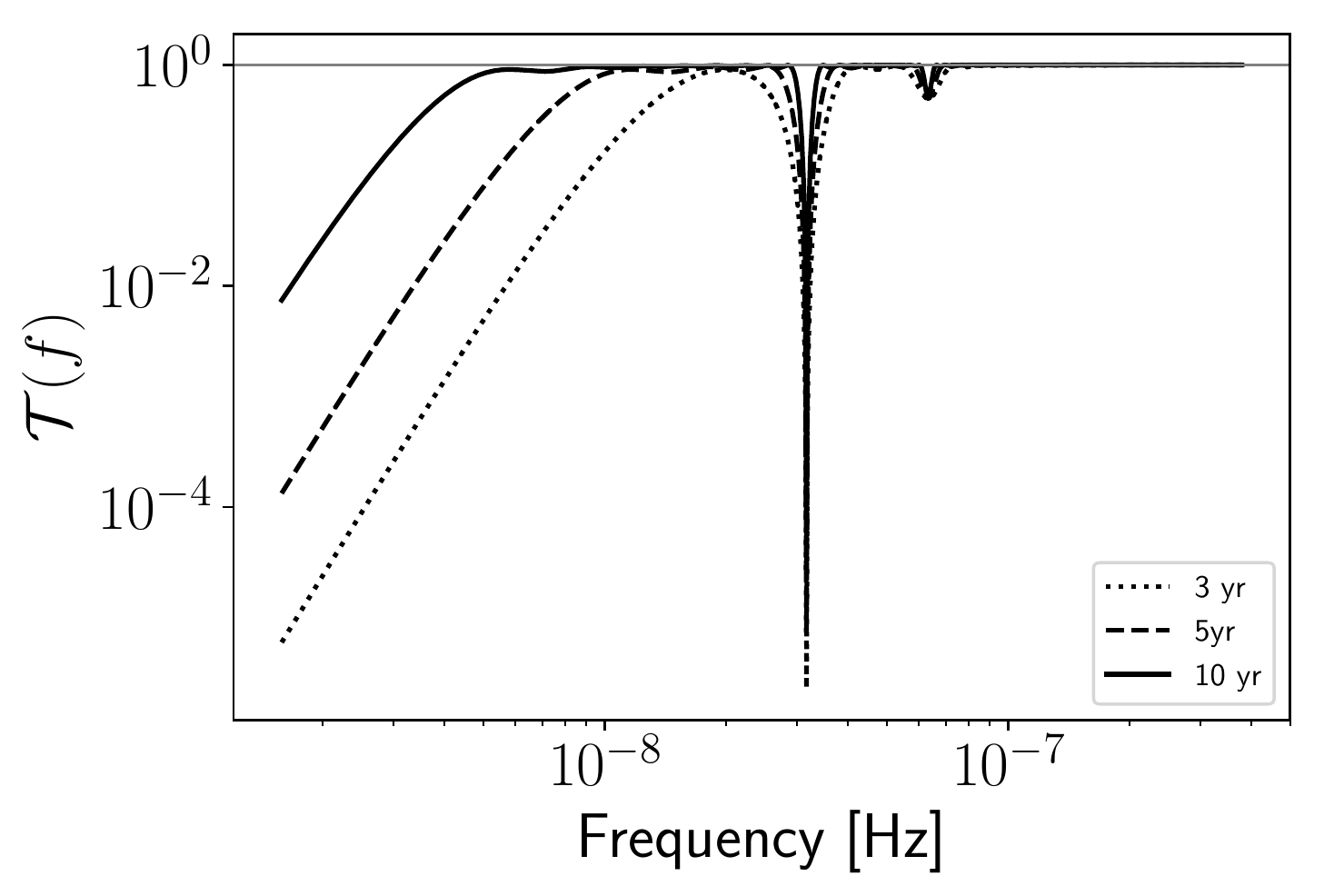}
\label{f:transmission_3_5_10}}
\caption{Transmission functions corresponding to a fit to a simple
timing model.
Panel (a): Absorption of power due to fitting to the quadractic
spin-down model, pulsar sky position, and distance to the pulsar
(parallax) are shown.
The blue vertical line corresponds to a frequency of $1/T$, 
where $T$ is the observation time.
Panel (b): Dependence of the transmission function on the 
duration of the observation.
The spikes become deeper and narrower, and the knee frequency
shifts to the left, as the observing time $T$ increases.}
\end{figure*}

The process of fitting to a timing model removes power
from the post-fit or TMM residuals.
This can be easily demonstrated by calculating the 
variance of the TMM residuals $r \equiv G^T\,\delta t$.
One finds
\be
\sigma^2_r = 
\int_0^\infty {\rm d}f\> {\cal T}(f)P(f)\,,
\label{e:sigma_r}
\ee
where $P(f)$ is the (one-sided) power spectral density of the 
original (pre-fit) timing residuals $\delta t$, and
\be
{\cal T}(f)\equiv
\frac{1}{N}\sum_{k,l}
(G G^T)_{kl} e^{i2\pi f(t_k-t_l)}\,.
\label{e:transmission}
\ee
The function ${\cal T}(f)$ has the interpretation of
a {\em transmission function}, selectively 
removing power associated with the timing model fit.
A plot of ${\cal T}(f)$ for a simple timing model
consisting of quadratic spin-down (i.e., fitting to the 
phase offset, spin period, and period derivative of the pulsar),
the pulsar's sky position, and the distance to the pulsar 
is shown in Figure~\ref{f:transmission}.
Note that fitting to the sky position absorbs power at and around a 
frequency of 1/year, corresponding to the Earth's yearly 
orbital motion around the Sun.
Fitting to the pulsar distance absorbs power at a frequency
of 2/year, which corresponds to a parallax measurement.
The quadratic spin-down parameter fit acts as a high-pass filter, 
absorbing frequencies substantially below $1/T$, where
$T$ is the time span of the data.
The effect of the observing time on the shape of the transmission
function is shown in Figure~\ref{f:transmission_3_5_10}.

Pulsars in binaries famously have additional components to the timing
model that take into account the various Doppler shifts due to binary
motion and relativistic effects, if the line-of-sight passes by the
companion (Shapiro delay) or if the binary is in a tight enough orbit 
to observe the loss of power due to GWs \cite{Taylor:1982zz}. These
components of the timing model have a minimal effect on sensitivity
curves for GWs as the frequencies in question are much
higher than those of the sources for which PTAs are searching. We
do not include these components when simulating pulsar design
matrices, but we will see the (mostly subtle) changes they make 
when looking at the design matrices of real pulsar data.  

Finally, we note that one can also calculate an analogous
transmission function associated with the post-fit timing
residuals $\delta t^{\rm post} \equiv R\,\delta t$.
One finds 
\be
\sigma^2_{\rm post} = \int_0^\infty
{\rm d}f\> {\cal T}_R(f) P(f)\,,
\ee
where
\be
{\cal T}_R(f) \equiv
\frac{1}{N}\sum_{k,l} R_{kl}\, e^{i2\pi f(t_k-t_l)}\,.
\label{e:transmission_R}
\ee
This $R$-matrix  transmission function was originally described
in \cite{Blandford:1984a}, although from a slightly different 
perspective.
In Appendix~\ref{s:appendix}, we cast the 
approach of \cite{Blandford:1984a}
into the more modern $R$-matrix notation.

\subsection{Inverse-noise-weighted transmission function}
\label{s:Ninvff'}

\begin{figure*}[h!tbp]
\centering
\subfigure[]{\includegraphics[width=0.48\textwidth]{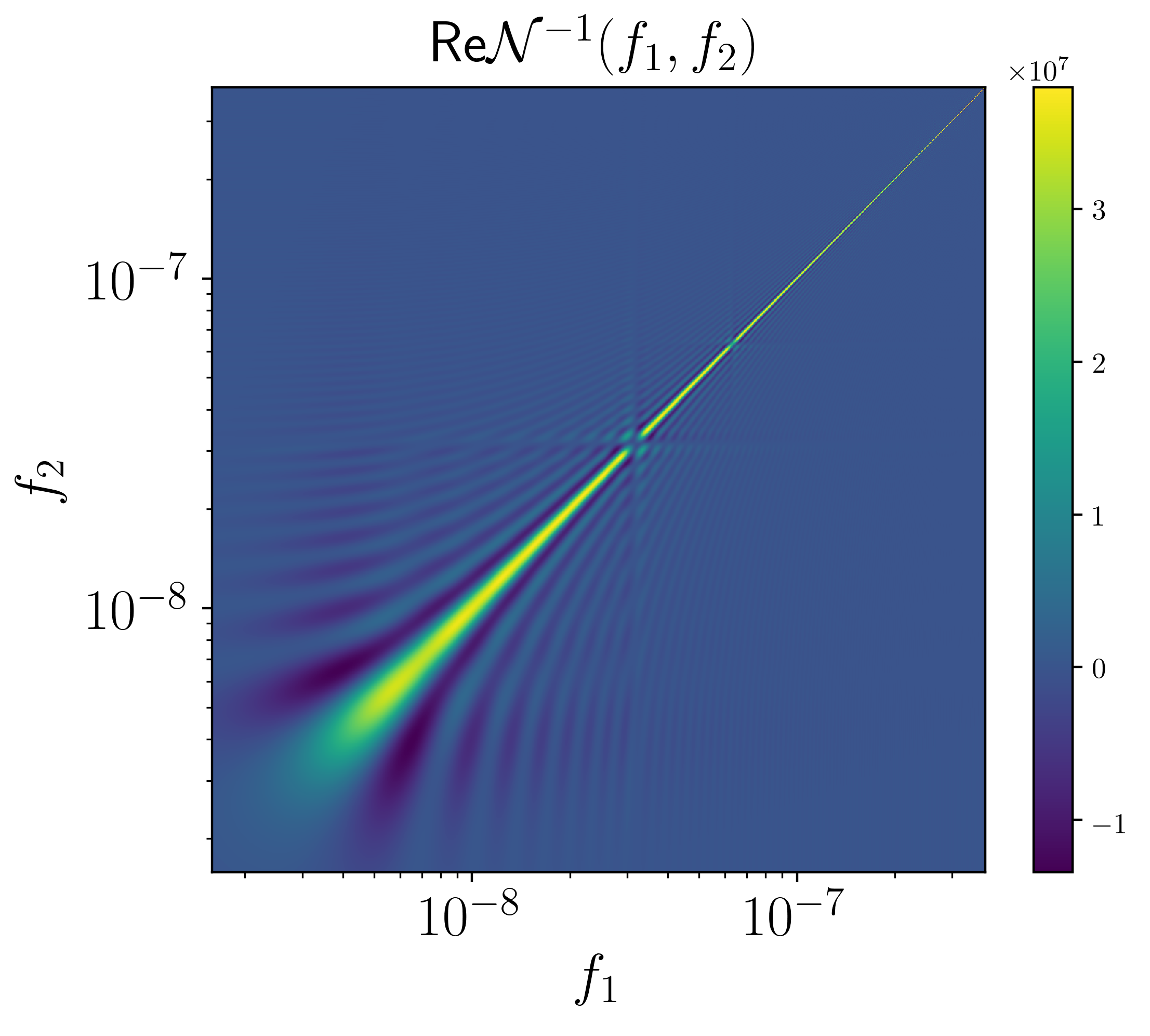}
\label{f:Ninvff'_a}}
\subfigure[]{\includegraphics[width=0.48\textwidth]{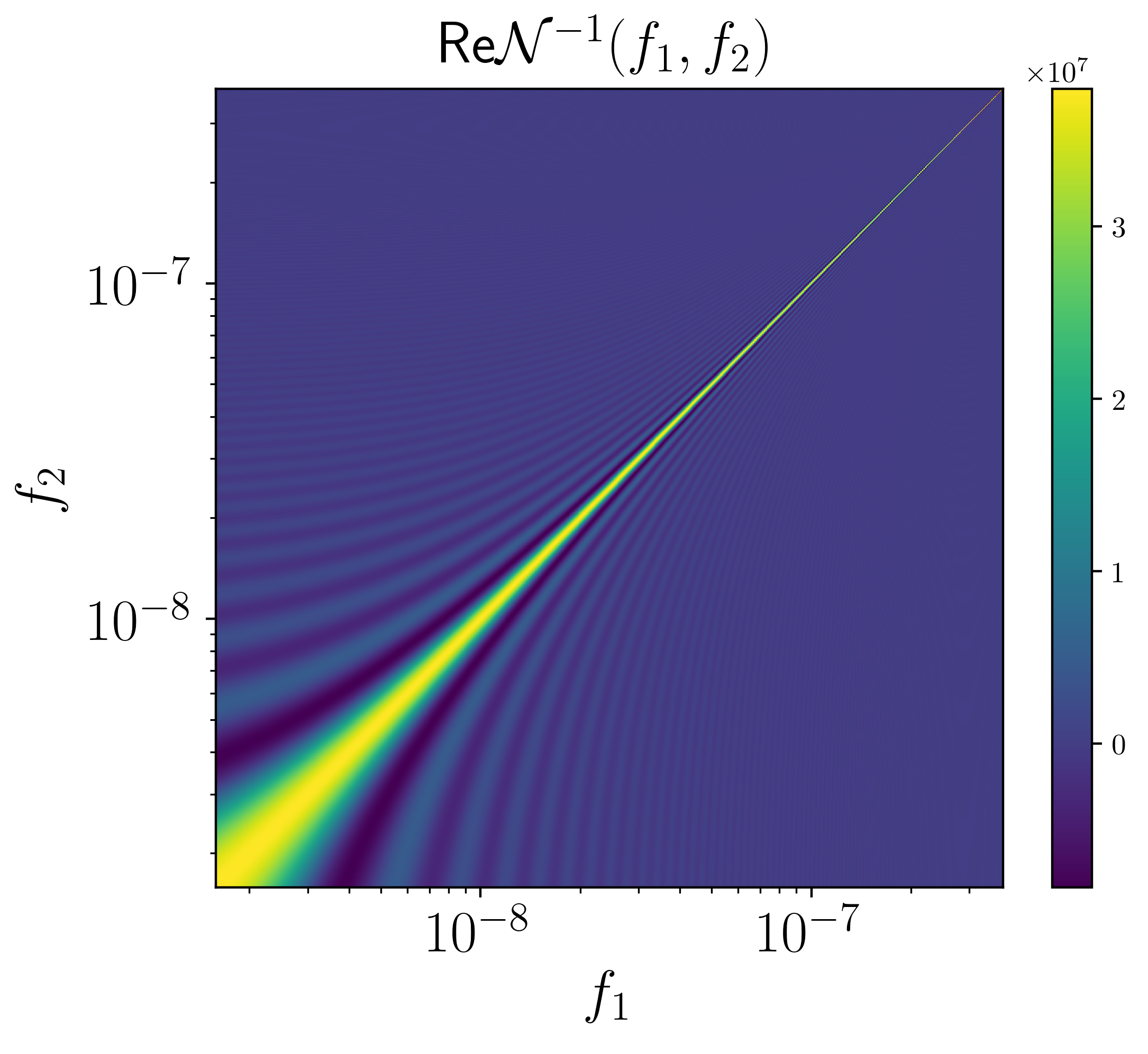}
\label{f:Ninvff'_b}}
\caption{Two-dimensional 
plot of the real part of the function ${\cal N}^{-1}(f_1,f_2)$ for $f_1,f_2>0$
plotted on log-scale axes.
Panel (a): ${\rm Re}[{\cal N}^{-1}(f_1,f_2)]$ for white noise 
($C$ is proportional to the identity matrix)
and a fit to the simple quadratic spin-down timing model described in the main text.
The small amplitude in the bottom-left hand corner of the plot
is due to the absorption of power by the timing model fit at and 
below $1/T$.
There is also suppression at $f_1=f_2=1/{\rm yr}$ and $f_1=f_2=2/{\rm yr}$.
Panel (b): For comparison, a two-dimensional plot of 
${\rm Re}[{\cal N}^{-1}(f_1,f_2)]$ for white noise,
but without performing a timing model 
fit (so $G$ is proportional to the identity matrix).}
\label{f:Ninvff'}
\end{figure*}

\begin{figure*}[h!tbp]
\centering
\subfigure[]{\includegraphics[width=0.49\textwidth]{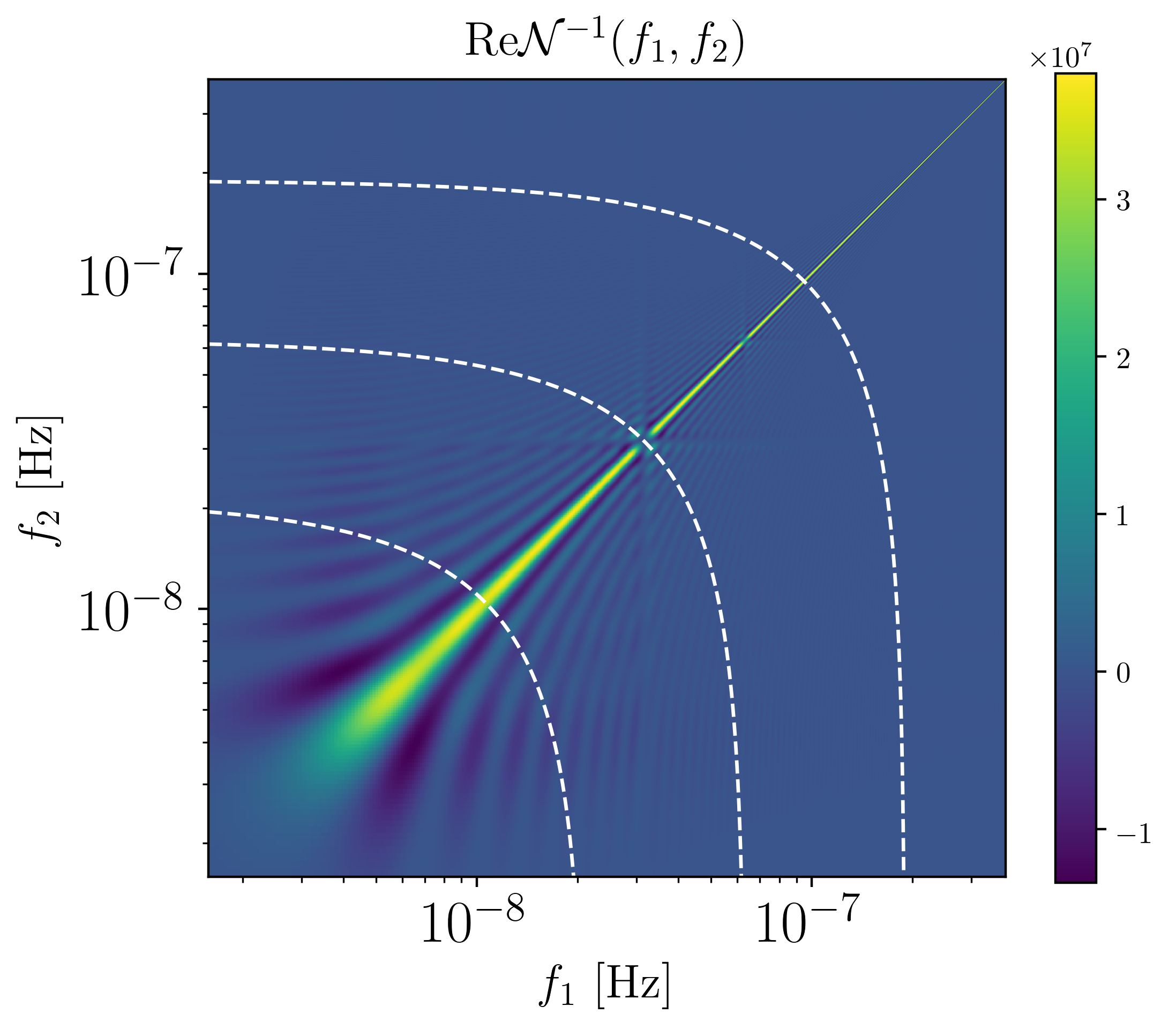}}
\subfigure[]{\includegraphics[width=0.49\textwidth]{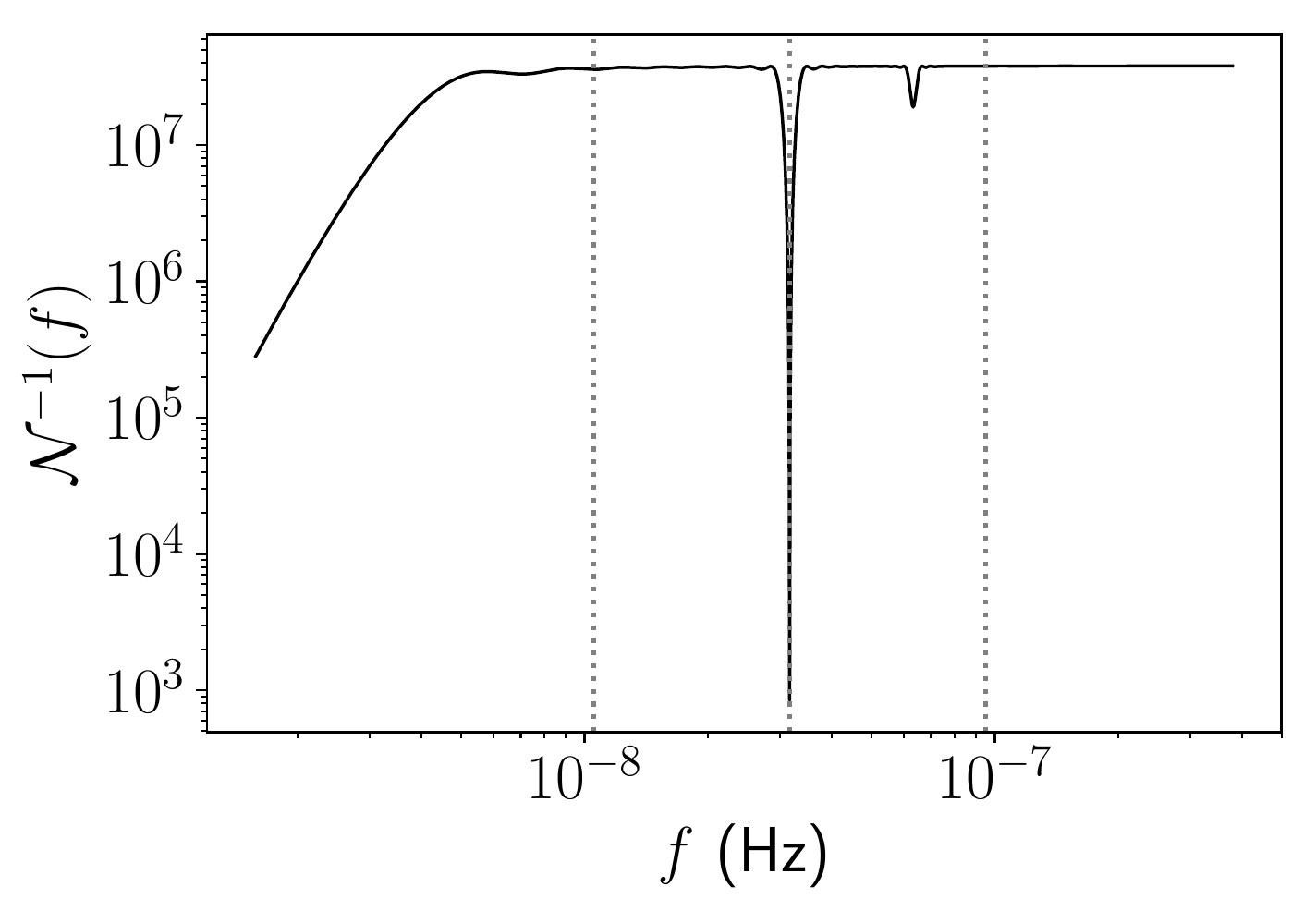}}
\subfigure[]{\includegraphics[width=0.32\textwidth]{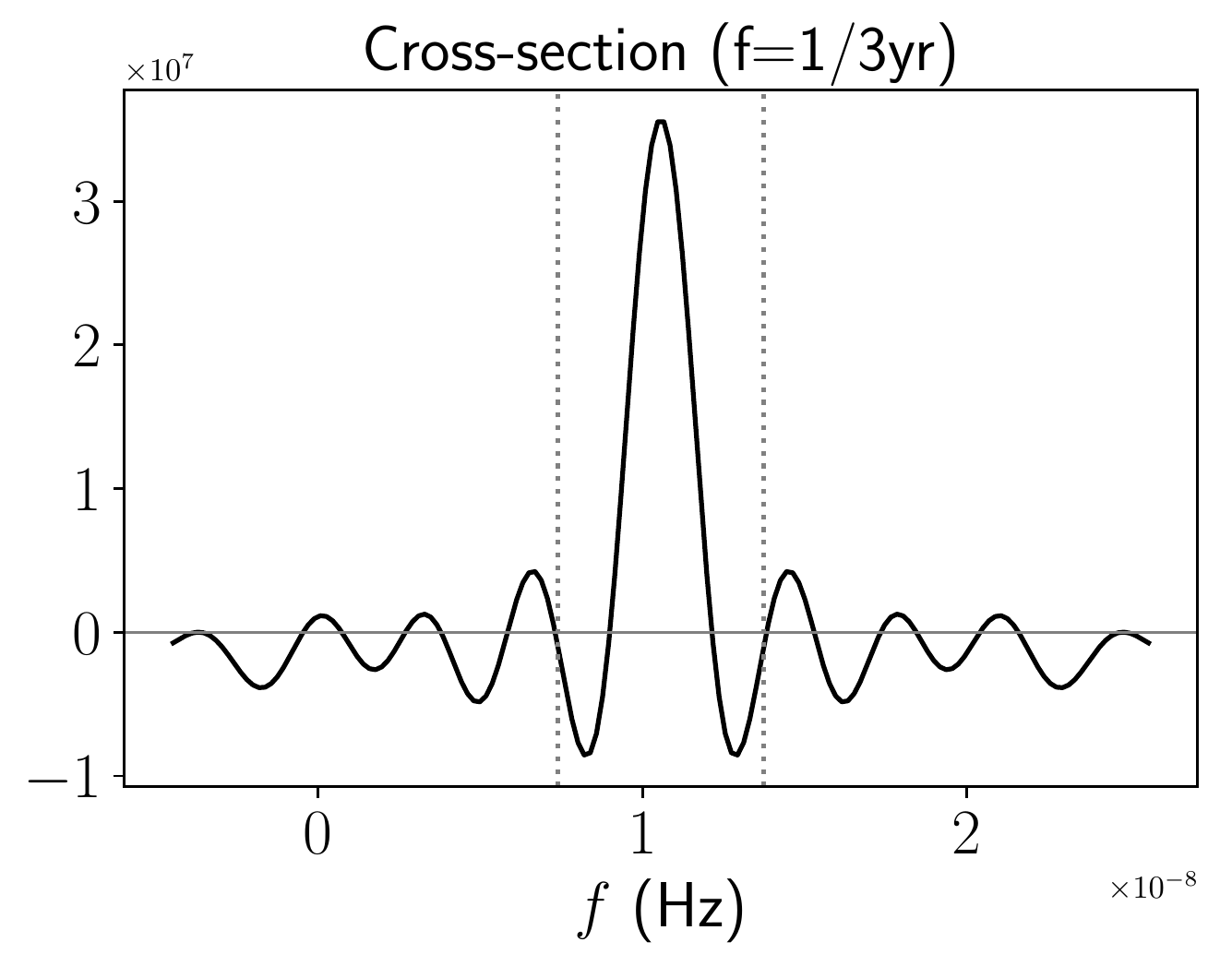}}
\subfigure[]{\includegraphics[width=0.32\textwidth]{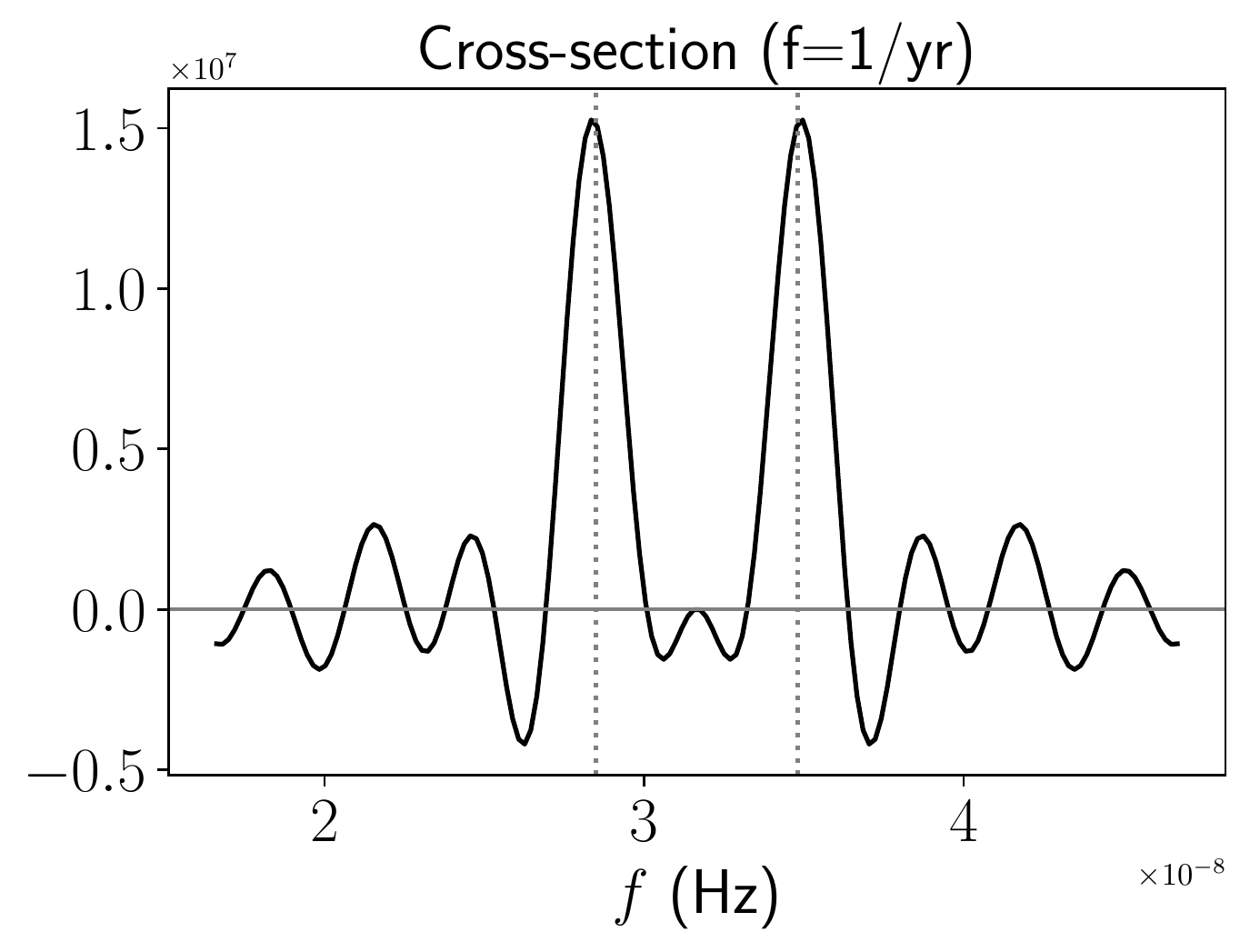}}
\subfigure[]{\includegraphics[width=0.32\textwidth]{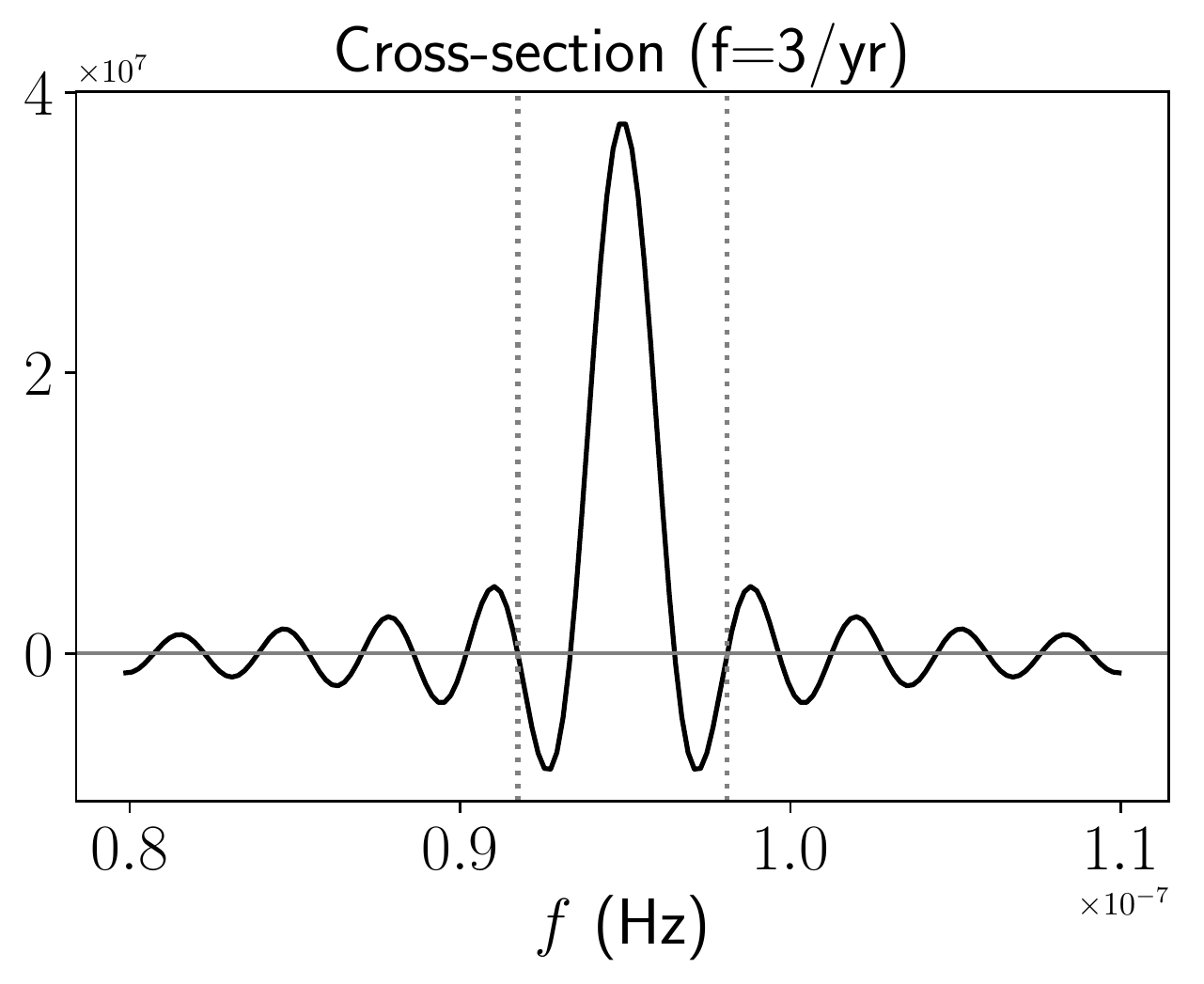}}
\caption{Diagonal and off-diagonal cross-sections of 
${\cal N}^{-1}(f_1,f_2)$.
Panel (a): ${\rm Re}[{\cal N}^{-1}(f_1,f_2)]$ from Figure~\ref{f:Ninvff'_a}
with off-diagonal cross-sections shown by white-dotted lines.
Panel (b): Diagonal component ${\cal N}^{-1}(f)$ 
(the dotted vertical lines show the frequencies of the off-diagonal 
cross-sections).
Panels (c)-(e): Real part of the off-diagonal cross sections of 
${\cal N}^{-1}(f_1,f_2)$ for $f=1/(3~{\rm yr})$, $f=1/{\rm yr}$, and 
$f=3/{\rm yr}$.
Away from 1/yr, the off-diagonal cross-sections are 
proportional to Dirichlet sinc functions (the dotted vertical 
lines indicate off-sets of $\pm 1/T$).}
\label{f:Ninvff'_cross_sections}
\end{figure*}

\begin{figure*}[h!tbp]
\centering
\includegraphics[width=\textwidth]{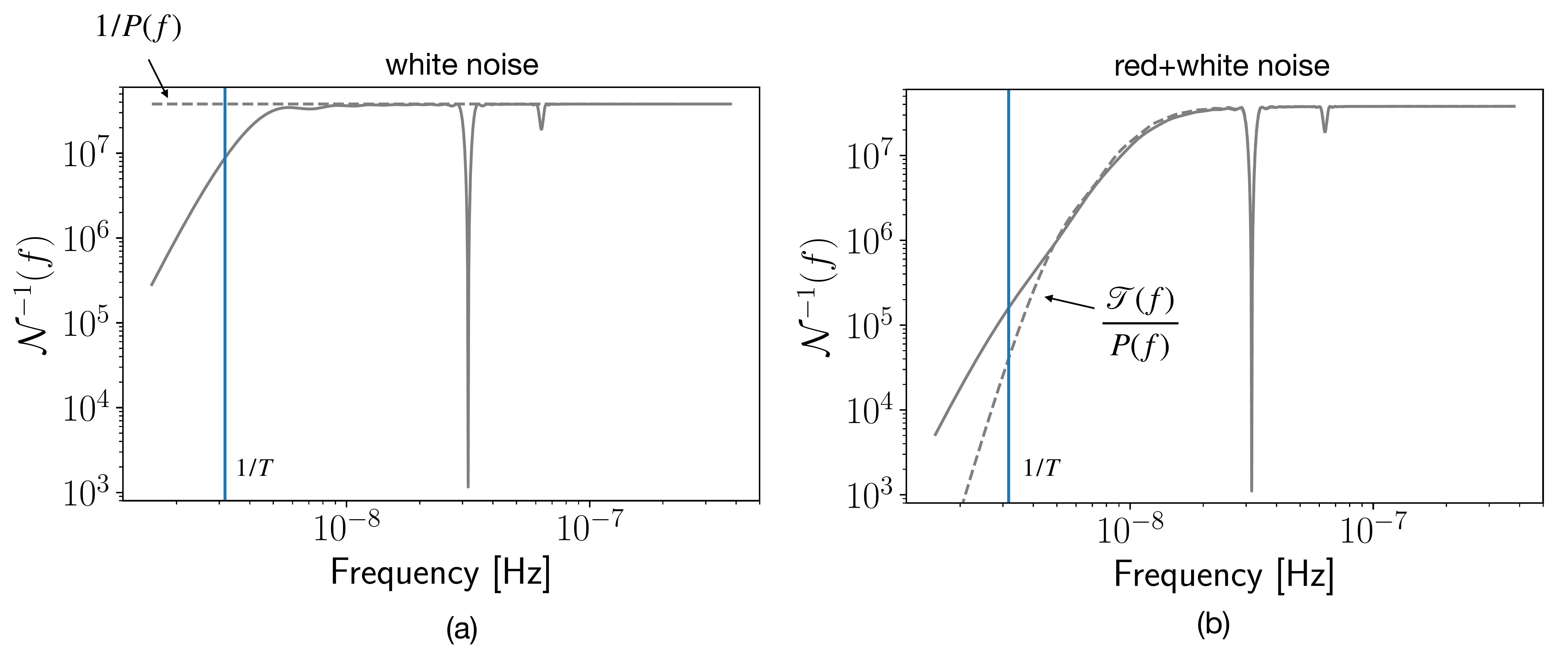}
\caption{Plots of the inverse-noise-weighted transmission function
${\cal N}^{-1}(f)$ for the simple quadratic spin-down model 
described in the main text, and for white noise (a) and red+white noise (b).
Panel (a): For white noise, the amplitude of ${\cal N}^{-1}(f)$ is 
set by the constant value of $1/P(f)$ indicated by the horizontal dashed line.
Panel (b): The curved dashed line is a plot of ${\cal T}(f)/P(f)$,
which is an approximation to ${\cal N}^{-1}(f)$ for $P(f)$ consisting 
of red+white noise.}
\label{f:Ninv_annotated}
\end{figure*}

It turns out that there is another way of obtaining a quantity 
that behaves like a transmission function by working directly with 
the TMM likelihood \eqref{e:likelihood_TMM}.
The argument of the exponential can be written as $-\frac{1}{2}\chi^2$, 
where 
\be
\chi^2
\equiv (\delta t-h(\theta))^TG(G^TC G)^{-1}G^T(\delta t -h(\theta))\,.
\ee
If we write this in the Fourier domain by substituting
\be
h_k(\theta) 
\equiv h(t_k;\theta) = \int_{-f_{\rm Nyq}}^{f_{\rm Nyq}} {\rm d}f\> 
\tilde h(f;\theta)\,e^{i2\pi f t_k}\,,
\ee
where $t_k\equiv k\Delta t$ and $f_{\rm Nyq} \equiv 1/(2\Delta t)$, we find
\begin{multline}
\chi^2 =2T\int_{-f_{\rm Nyq}}^{f_{\rm Nyq}}{\rm d}f\int_{-f_{\rm Nyq}}^{f_{\rm Nyq}}{\rm d}f'\>
(\widetilde{\delta t}(f)-\tilde h(f;\theta))
\\
\times {\cal N}^{-1}(f,f') (\widetilde{\delta t}{}^*(f')-\tilde h^*(f';\theta))\,,
\end{multline}
where
\begin{multline}
{\cal N}^{-1}(f,f')
\\
\equiv \frac{1}{2T}
\sum_{k,l} e^{i2\pi ft_k}\,
[G(G^TC G)^{-1}G^T]_{kl}\,e^{-i2\pi f't_l}\,.
\label{e:Ninvff'}
\end{multline}
The quantity ${\cal N}^{-1}(f,f')$ 
is a function of two frequencies, $(f,f')$, but it 
turns out to be {\em diagonally-dominated}, with the majority  
of its support on the diagonal $f=f'$, as shown in 
Figure~\ref{f:Ninvff'_a}.
(The broadening of the diagonal band at low frequencies is an artefact
of using log-scale axes for the frequencies.)
The diagonal component
\be
{\cal N}^{-1}(f)\equiv 
\frac{1}{2T}\sum_{k,l}
[G(G^TC G)^{-1}G^T]_{kl}\,e^{i2\pi f(t_k-t_l)}\,,
\label{e:Ninvf}
\ee
and three off-diagonal cross-sections of ${\cal N}^{-1}(f,f')$ 
are shown in Figure~\ref{f:Ninvff'_cross_sections}.
(The fact that the off-diagonal cross-sections are curved in
panel (a) of Figure~\ref{f:Ninvff'_cross_sections} is again
due to using log-scale axes for the frequencies.)
A few remarks are in order:

(i) For this particular example, the diagonal component 
${\cal N}^{-1}(f)$ is {\em identical} in shape with 
the transmission function 
${\cal T}(f)$ shown in Figure~\ref{f:transmission}.
The amplitude of ${\cal N}^{-1}(f)$ differs 
from ${\cal T}(f)$ by a constant
factor $1/P(f) = 1/(2\sigma^2\,\Delta t)$, corresponding to 
a white noise covariance matrix.%
\footnote{For our white noise simulations, we take
$P(f) = 2\sigma^2\,\Delta t$, with $\sigma=100~{\rm ns}$ 
and $\Delta t = {\rm yr}/20$.  These numerical values are
often chosen for pulsar timing simulations.}
Thus, for white noise
\be
{\cal N}^{-1}(f) = {{\cal T}(f)}/{P(f)}\,.
\ee
This is illustrated in Figure~\ref{f:Ninv_annotated}(a).
If we also include red noise in the noise covariance
matrix $C$ by taking
\begin{align}
&C_{ij} = \int_0^{f_{\rm Nyq}} 
{\rm d}f\>\cos[2\pi f(t_i-t_j)]\,P(f)\,,
\\
&P(f) = 2\sigma^2\,\Delta t + A f^{-\gamma}\,,\qquad  \gamma>0\,,
\end{align}
then the relationship between ${\cal N}^{-1}(f)$
and ${\cal T}(f)/P(f)$ is only approximate,
\be
{\cal N}^{-1}(f) \approx {{\cal T}(f)}/{P(f)}\,.
\ee
This is illustrated in Figure~\ref{f:Ninv_annotated}(b).

(ii) Away from the dip at 1/yr, where there is suppression
of power due to the timing model fit to the pulsar sky position, 
the off-diagonal cross-sections are proportional to 
Dirichlet sinc functions 
\be
{\cal D}_N((f-f')\Delta t)\equiv \frac{1}{N}\frac{\sin[N\pi(f-f')\Delta t]}
{\sin[\pi(f-f')\Delta t]}\,.
\label{e:dirichlet}
\ee
When multiplied by $T$, a Dirichlet sinc function can be 
thought of as 
finite-time approximation to the Dirac delta function---i.e.,
$\delta(f-f')\simeq T{\cal D}_N((f-f)'\Delta t)$.
Dirichlet sinc functions arise when taking the Fourier transform of 
a discretely-sampled rectangular window of duration 
$T = N\Delta t$, see e.g.,~\cite{Romano:2016dpx}.
This diagonally-dominated behavior is what you would expect 
for ${\cal N}^{-1}(f,f')$  if one had only 
Gaussian-stationary noise.
This is the case if one doesn't have to fit a 
timing model (Figure~\ref{f:Ninvff'_b}).
Then one can simply 
replace $G$ by the identity matrix, for which
\be
\begin{aligned}
{\cal N}^{-1}(f,f')
&=\frac{1}{2T}\sum_{k,l}
e^{i2\pi f t_k} [C^{-1}]_{kl}\,e^{-i2\pi f't_l}
\\
&\simeq P^{-1}(f)\,\delta_{ff'}\,.
\end{aligned}
\ee
The approximate equality in the above equation 
is a consequence of the Karhunen-Loeve theorem, 
which states that the discrete Fourier transform operation
defined by the unitary matrix $U_{jk} \equiv \frac{1}{N}e^{-i2\pi jk/N}$
approximately diagonalizes a stationary covariance matrix
in the limit that the observation time $T=N\,\Delta t$ 
is much larger than the correlation time of the noise.

(iii) Since fitting to a timing model introduces non-stationarities
into the TMM residuals~\cite{vanHaasteren:2012hj}, one cannot directly
appeal to the Karhunen-Loeve theorem for the general
expression \eqref{e:Ninvff'}.
One needs to explicitly check the validity of the 
diagonal approximation
for ${\cal N}^{-1}(f,f')$ as we have done in 
Figures~\ref{f:Ninvff'} and \ref{f:Ninvff'_cross_sections}.
We have also numerically computed the sum of 
${\cal N}^{-1}(f,f')$ over the full two-dimensional 
array of frequencies $(f,f')$ and compared that to the sum 
of ${\cal N}^{-1}(f,f')$ just along the diagonal $f=f'$.  
Even for the more challenging 
case of a red+white noise covariance matrix
(Figure~\ref{f:Ninv_annotated}(b))
and a fit to the our quadratic spin-down model, the 
two summations agree to within $\approx 6\%$.

%% file: manuscript/response.tex
\section{Timing residual response to gravitational waves}
\label{s:response}

To proceed further in our calculation of pulsar timing sensitivity
curves, we need to describe in more detail the timing residual
response of a pulsar to an incident GW.
We will consider both deterministic and stochastic sources
of GWs.
Interested readers should see \cite{Blandford:1984a, 2007PhDT........14D, vanHaasteren:2008yh, 2011MNRAS.418..561C, vanHaasteren:2012hj} for more details.

\subsection{Response to a single deterministic source}
\label{s:deterministic}

\begin{figure}[h!tbp]
\centering
\includegraphics[width=0.8\columnwidth]{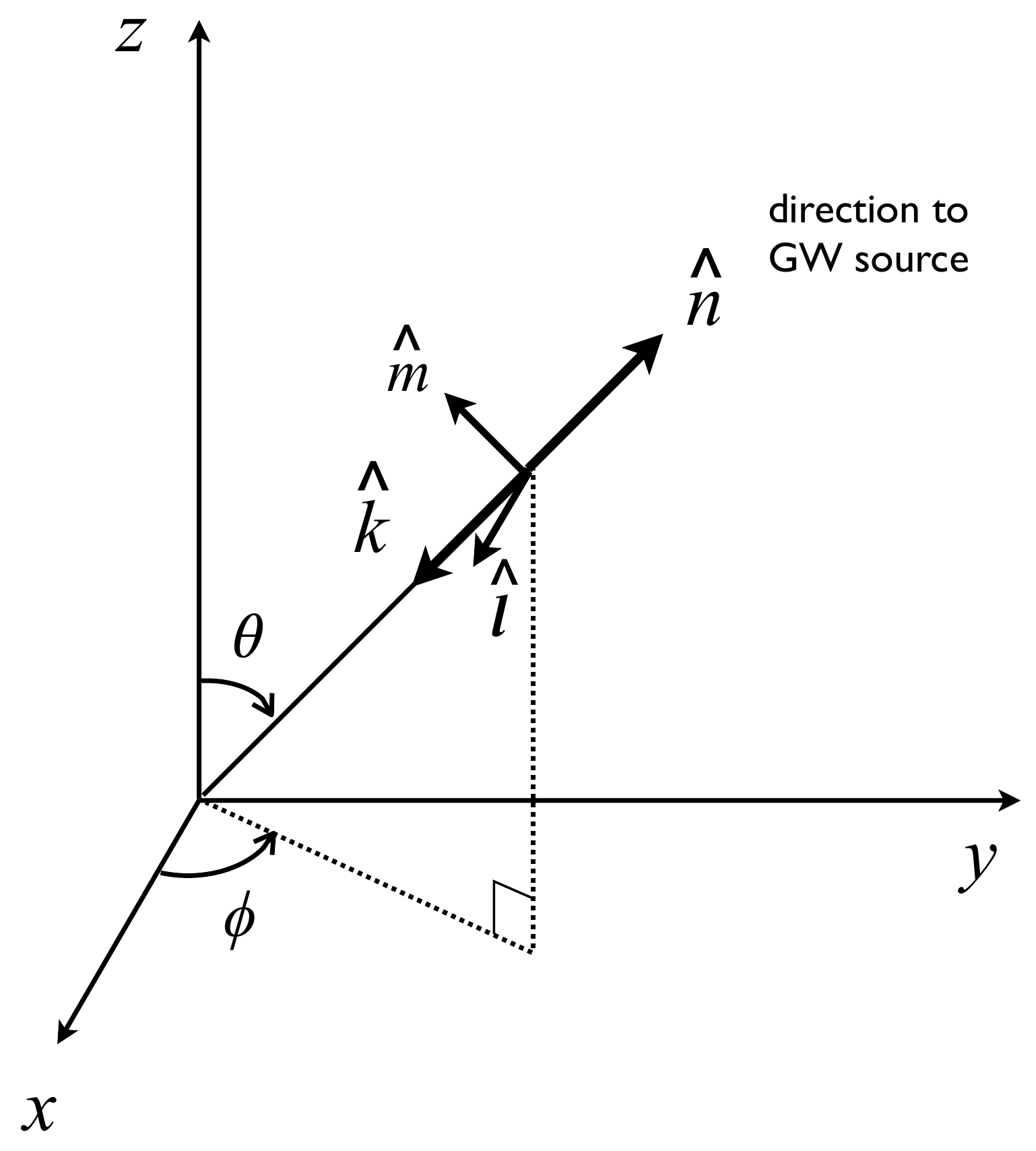}
\caption{Definition of the unit vectors $\hat k$, $\hat l$,
$\hat m$.
The direction of propagation of the GW,
$\hat k$, is opposite the direction to the source, $\hat n$.
The unit vectors $\hat l$, $\hat m$ are in the plane perpendicular
to $\hat k$, and point in directions of constant declination 
and right ascension, respectively.}
\label{f:plane_wave}
\end{figure}

\begin{figure*}[h!tbp]
\centering
\subfigure[]{\label{f:inclination_angle}
\includegraphics[width=0.25\textwidth]{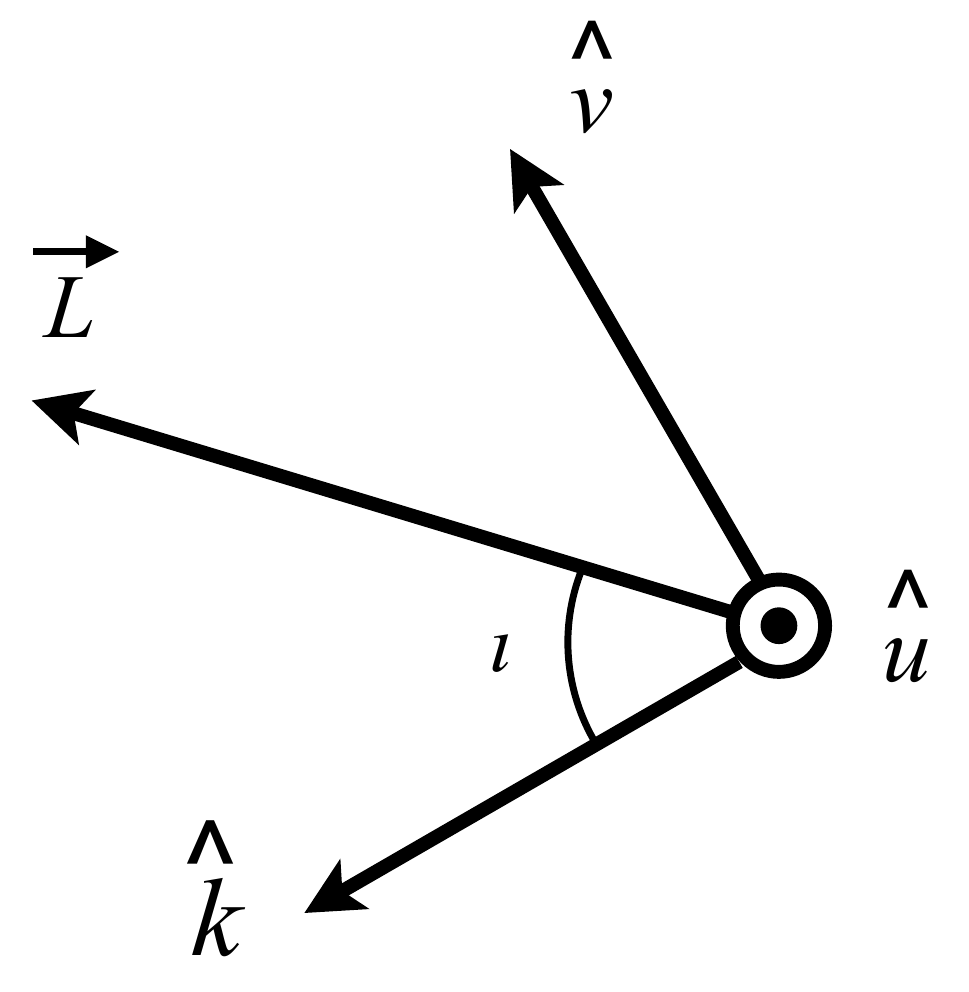}}
\subfigure[]{\label{f:polarization_angle}
\hspace{.5 true in}
\includegraphics[width=0.35\textwidth]{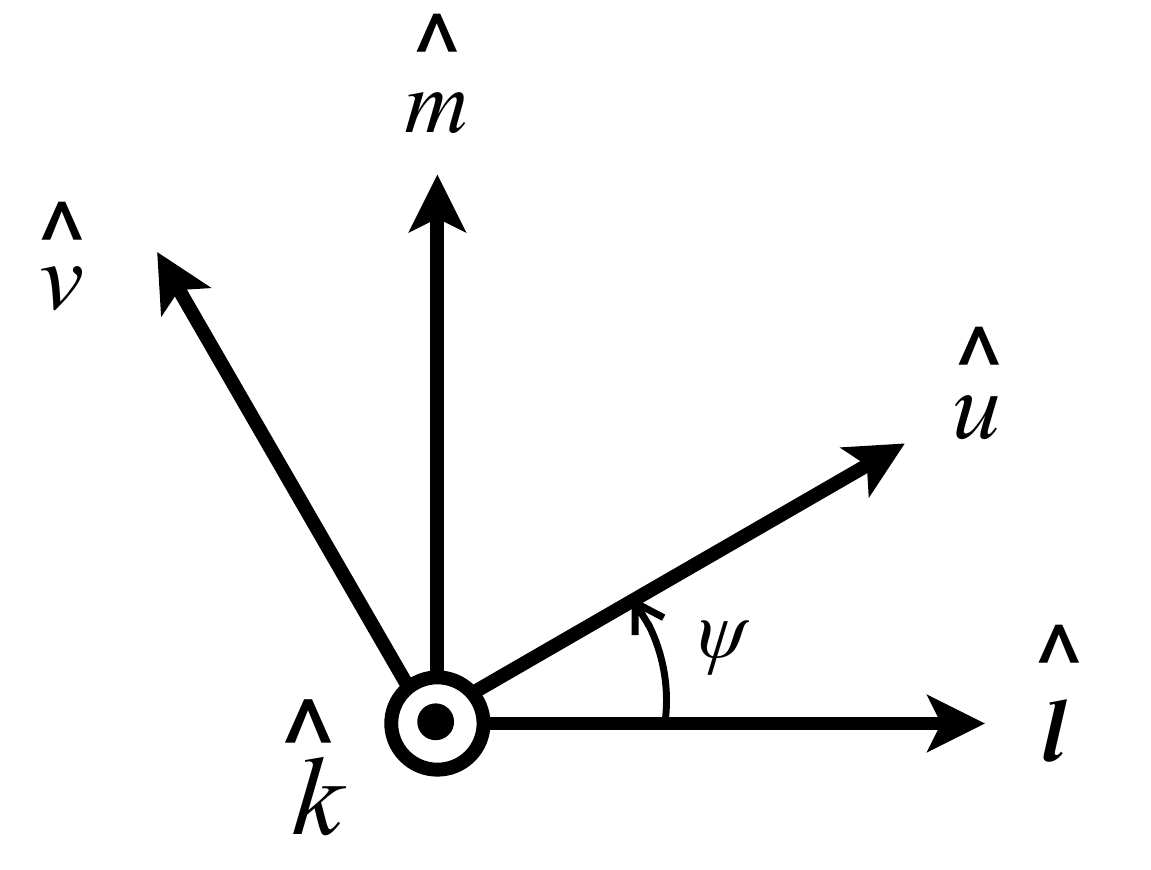}}
\caption{Relation between the unit vectors $\hat l$, $\hat m$ and
$\hat u$, $\hat v$.
Definition of: (a) inclination angle $\iota$, and
(b) polarization angle $\psi$.
Here $\vec L$ is the angular momentum vector and $\hat k$
is the propagation direction of the GW.
The vectors $\hat l$, $\hat m$ and $\hat u$, $\hat v$ 
are orthogonal unit vectors in the plane perpendicular to $\hat k$, 
defined by \eqref{e:klm} and \eqref{e:u,v}, respectively.}
\label{f:inclination}
\end{figure*}

We will start by writing down the metric perturbations
$h_{ab}(t,\vec x)$ for a single deterministic source emitting
plane GWs in the direction $\hat k$ 
(Figure~\ref{f:plane_wave}). To do this we introduce
two coordinate frames: one associated with the solar system 
barycenter (SSB) and the other associated with the propagation 
of the GW.
We will assume that the source has a symmetry axis 
(e.g., the direction of the orbital angular momentum vector
$\vec L$ for a binary system), and that the symmetry axis 
makes an angle $\iota$ with respect to the line of sight 
$\hat k$ from the GW source to the solar system barycenter,
and an angle $\psi$ with respect to the vector $\hat l$ when 
projected onto the plane perpendicular to $\hat k$ 
(Figure~\ref{f:inclination}).
The vectors $\hat k$, $\hat l$, $\hat m$ are defined in the 
solar system barycenter frame by
\be
\begin{aligned}
\hat{k} &= (-\sin\theta \, \cos\phi, -\sin\theta\,\sin\phi,-\cos\theta)
\equiv -\hat r\,,
\\
\hat{l} &= (\sin\phi,-\cos\phi, 0)\equiv -\hat\phi\,,
\\ 
\hat{m} &= (-\cos\theta\, \cos\phi, -\cos\theta\, \sin\phi, \sin\theta)
\equiv -\hat\theta\,,
\label{e:klm}
\end{aligned}
\ee
where $(\theta,\phi)$ are the standard polar and azimuthal
angles on the 2-sphere in equatorial coordinates, and the 
origin of coordinates is at the solar system  barycenter.
The right ascension $\alpha$ and declination $\delta$ of a
source are given in terms of $\theta$ and $\phi$ by
$\alpha=\phi$ and $\delta = \pi/2-\theta$.

The angles $\iota$ and $\psi$ are the {\em inclination} 
and {\em polarization} angles of the source, respectively.
They can be written in terms of the unit vectors $\hat k$,
$\hat l$, $\hat L \equiv \vec L/|\vec L|$, and $\hat u$
via:
\be
\cos\iota \equiv \hat k\cdot\hat L\,,
\qquad
\cos\psi\equiv \hat u\cdot\hat l\,,
\label{e:polarization}
\ee
where
\be
\hat u\equiv \frac{\hat L\times \hat k}{|\hat L\times\hat k|}\,,
\qquad
\hat v\equiv \hat{k}\times\hat{u}\,,
\label{e:u,v}
\ee
are two orthogonal unit vectors in the plane perpendicular 
to $\hat k$ (Figure~\ref{f:inclination}).
Note that $\iota=0$ or $\pi$ corresponds to the orbital
plane being seen face-on or face-off; 
$\iota=\pi/2$ or $3\pi/2$ corresponds to seing the 
orbital plane edge on.
The unit vectors $\hat u$, $\hat v$ are related to
$\hat l$, $\hat m$ by a rotation around $\hat k$
through the angle $\psi$ as shown in Figure~\ref{f:polarization_angle}.

From $\hat u$ and $\hat v$, we can construct a preferred
set of polarization tensors:
\be
\begin{aligned}
\epsilon^{+}_{ab}(\hat k,\psi) 
&\equiv \hat{u}_a \hat{u}_b - \hat{v}_a \hat{v}_b\,, 
\\ 
\epsilon^{\times}_{ab}(\hat k,\psi)
&\equiv \hat{u}_a \hat{v}_b + \hat{v}_a\hat{u}_b\,.
\label{e:eps+,epsx}
\end{aligned}
\ee
Using these polarization tensors, we can expand the metric perturbations:
\begin{multline}
h_{ab}(t,\vec x)
= h_+(t-\hat k\cdot\vec x/c; \iota)\epsilon^+_{ab}(\hat k,\psi) 
\\
+ h_\times(t-\hat k\cdot\vec x/c; \iota)\epsilon^\times_{ab}(\hat k,\psi)
\label{e:hab(t,x)_det_time}
\end{multline}
or, equivalently,
\begin{multline}
h_{ab}(t,\vec x)
=\int_{-\infty}^\infty{\rm d}f\>
\left[\tilde h_+(f;\iota) \epsilon^+_{ab}(\hat k,\psi) 
\right.
\\
\left.
+\tilde h_\times(f;\iota) \epsilon^\times_{ab}(\hat k,\psi)\right]\,
e^{i2\pi f(t-\hat k\cdot\vec x/c)}\,,
\label{e:hab(t,x)_det_freq}
\end{multline}
where $\tilde h_{+,\times}(f;\iota)$ are the Fourier transforms 
of $h_{+,\times}(t;\iota)$.
The timing residual response of a pulsar to such a deterministic 
GW is then~\cite{Detweiler:1979wn}:
\be
h(t;\hat k, \iota, \psi) = \int_{-\infty}^\infty {\rm d}f\>
\tilde h(f;\hat k,\iota, \psi)\,e^{i2\pi ft}\,,
\ee
where
\begin{multline}
\tilde h(f;\hat k,\iota,\psi) 
= R^+(f,\hat k,\psi)\tilde h_+(f;\iota)
\\
+ R^\times(f,\hat k,\psi)\tilde h_\times(f;\iota)\,,
\label{e:htildeI_det}
\end{multline}
with
\begin{multline}
R^{+,\times}(f,\hat k,\psi)\equiv
\frac{1}{i 2\pi f}
\frac{1}{2}\frac{\hat{p}^a \hat{p}^b}{1+\hat{p}\cdot \hat k}
\epsilon^{+,\times}_{ab}(\hat k,\psi) 
\\
\times
\left(1-e^{-i 2\pi f D(1+\hat k\cdot\hat p)/c}\right)\,.
\label{e:R+x}
\end{multline}
Here $\hat p^a$ is a unit vector pointing from the solar system
barycenter to the pulsar, and $D$ is the distance to the pulsar.
The function $R^{+,\times}(f,\hat k, \psi)$ is the timing
residual response function of a pulsar to a monochromatic plane GW 
propagating in direction $\hat k$, with frequency $f$, 
polarization $+,\times$, and polarization angle $\psi$.
The two terms in the response function 
are  called the `Earth term' and `pulsar term', respectively, 
since they involve sampling the GW phase at Earth and at
the location of the pulsar, a distance $D$ away from Earth.
The factor of $1/(i2\pi f)$ comes from the fact that we are 
working with timing residuals, as opposed to Doppler shifts in 
the pulse frequency.

For the analyses that we will do in this paper, 
we will typically ignore the pulsar-term contribution to 
the timing residual response to GWs, 
as this term will not contribute to the
cross-power when correlating the signal associated with 
distinct pulsars.
(The separation between pulsars ($\sim {\rm kpc}$) is much 
greater the wavelengths of the GWs that we are sensitive to,
which are of order $\lesssim 10~{\rm lyr}$.)
There is a contribution, however, to the auto-correlated 
power for a {\em single} pulsar, which comes from the 
exponential part of $|R^{+,\times}(f,\hat k,\psi)|^2$:
\begin{multline}
\left|1-e^{-i2\pi fD(1+\hat k\cdot\hat p)/c}\right|^2
\\
=2\left[1-\cos(2\pi fD(1+\hat k\cdot\hat p)/c)\right]
\simeq 2\,,
\label{e:pulsar-pulsar}
\end{multline}
where we have ignored the cosine term since 
it is a rapidly-oscillating function of the 
GW propagation direction $\hat k$, and hence
does not contribute significantly when summed 
over the sky.
The value `2' corresponds to the sum of the Earth-Earth and 
pulsar-pulsar auto-correlation terms.

\subsubsection{Circular binaries}
\label{s:circular_binaries}

To proceed further, we need to specify the form of 
$h_{+,\times}(t;\iota)$ or its Fourier transform 
$\tilde h_{+,\times}(f;\iota)$.
For example, for a circular binary
\be
\begin{aligned}
h_+(t;\iota)
&= h_0(t) \left(\frac{1+\cos^2\iota}{2}\right)\,\cos 2\Phi(t)\,,
\\
h_\times(t;\iota)
&= h_0(t) \,\cos\iota\,\sin 2\Phi(t)\,,
\label{e:h+x(t)}
\end{aligned}
\ee
where $\Phi(t)$ is the orbital phase and $h_0(t)$ 
is a dimensionless amplitude given by 
\be
h_0(t) =\frac{4c}{D_L}
\left(\frac{G\mathcal{M}_{\rm c}}{c^3}\right)^{5/3}
\omega(t)^{2/3}\,.
\label{e:h0(t)}
\ee
Here $D_L$ is the luminosity distance to the source,
$\mathcal{M}_{\rm c} \equiv(m_1 m_2)^{3/5}/(m_1+m_2)^{1/5}$
is the {\em chirp mass} of the binary system, and
$\omega(t)$ is the instantaneous orbital angular 
frequency, $\Phi(t) = \int^t{\rm d}t'\>\omega(t')$.
For an evolving binary system
\be
\frac{{\rm d}\omega}{{\rm d}t}
= \frac{96}{5}\left(\frac{G\mathcal{M}_c}{c^3}\right)^{5/3}
\omega(t)^{11/3}\,,
\ee
which is a consequence of energy balance between the 
radiated power in GWs
and the orbital energy lost by the binary system.
The instantaneous GW frequency $f(t)$ is related to the 
orbital frequency $\omega(t)$ via $\omega(t) =\pi f(t)$. 

The above differential equation for $\omega(t)$ (or, 
equivalently, for $f(t)$) can be integrated to yield 
\be
f(\tau) = 
\frac{1}{\pi} 
\left(\frac{G\mathcal{M}_c}{c^3}\right)^{-5/8}
\left(\frac{5}{256}\frac{1}{\tau}\right)^{3/8}\,,
\label{e:f(tau)}
\ee
where $\tau\equiv t_{\rm col}-t$ is the time to 
coalescence.
Inverting \eqref{e:f(tau)}, we obtain
\be
\tau = \frac{5}{256}
\left(\frac{G\mathcal{M}_c}{c^3}\right)^{-5/3}
\left(\pi f\right)^{-8/3}\,,
\ee
which is the time to coalescence for a binary 
system currently having GW frequency $f$.
Note that for a SMBH binary with $10^9$ solar-mass 
BHs (which is the primary
source for PTAs) and GW frequency $f=8~{\rm nHz}$ 
(which is one of the most sensitive frequencies 
for the current decade-long PTA searches), the
time to coalescence is
$\tau\sim 10^5~{\rm yr}$, which is four orders of 
magnitude larger than a decade-long observation 
$T=10~{\rm yr}$.
Over the course of the observation the change in 
the GW frequency for the above SMBH binary is
\be
\Delta f 
\simeq
\frac{1}{\pi} 
\left(\frac{G\mathcal{M}_c}{c^3}\right)^{-5/8}
\left(\frac{5}{256}\frac{1}{\tau}\right)^{3/8}
\,\frac{3}{8}\frac{T}{\tau}
\approx 3\times 10^{-13}~{\rm Hz}\,,
\ee
which is four orders of magnitude smaller than the
frequency bin width $1/T$, set by the total
observation time $T$.
Thus, for the purposes of this paper, we will take 
our deterministic source to be a {\em monochromatic}
binary with $f(\tau)=f_0={\rm const}$.

With this simplification, equations \eqref{e:h+x(t)} 
and \eqref{e:h0(t)} become
\be
\begin{aligned}
h_+(t;\iota, \phi_0)
&= h_0 \left(\frac{1+\cos^2\iota}{2}\right)\,\cos(2\pi f_0 t + \phi_0)\,,
\\
h_\times(t;\iota, \phi_0)
&= h_0 \cos\iota\,\sin(2\pi f_0 t + \phi_0)\,,
\label{e:h+x(t)_f0}
\end{aligned}
\ee
where $\phi_0$ is the initial phase and $h_0$ is the 
(constant) strain amplitude
\be
h_0 =\frac{4c}{D_L}
\left(\frac{G\mathcal{M}_{\rm c}}{c^3}\right)^{5/3}
(\pi f_0)^{2/3}\,.
\label{e:h0}
\ee
The Fourier transforms of $h_{+,\times}(t;\iota,\phi_0)$ 
are then
\be
\begin{aligned}
\tilde h_+(f;\iota,\phi_0)
&= h_0 \left(\frac{1+\cos^2\iota}{2}\right)
\\
&\qquad
\times \frac{1}{2}\left[e^{i\phi_0}\delta(f-f_0) 
+ e^{-i\phi_0}\delta(f+f_0)\right]\,,
\\
\tilde h_\times(f;\iota,\phi_0)
&= h_0 \cos\iota\,
\\
&\qquad
\times \frac{1}{2i}\left[e^{i\phi_0}\delta(f-f_0) 
- e^{-i\phi_0}\delta(f+f_0)\right]\,.
\label{e:h+x(f)_f0}
\end{aligned}
\ee
But since the signals are observed for only a finite 
duration, the Dirac delta functions $\delta(f\mp f_0)$ 
should be replaced by their finite-time equivalents
$\delta_{T}(f\mp f_0)$ defined by
\be
\delta_{T}(f) 
\equiv \int_{-T/2}^{T/2}{\rm d}t\> e^{-i2\pi ft}
= \frac{\sin(\pi fT)}{\pi f}
\equiv T\,{\rm sinc}(\pi f T)\,,
\ee
where $T$ is the observation time for the pulsar.
If one wants to also include the discreteness $\Delta t$ of 
the time-series data, then the Dirac delta 
functions should be replaced by Dirichlet sinc functions, 
$T\mathcal{D}_{N}[(f\mp f_0)\Delta t]$ (see \eqref{e:dirichlet}).
It turns out that the 
final (approximate) expressions that we obtain, 
cf.~\eqref{e:|htildeI|^2_inc} and \eqref{e:|htildeI|^2_inc,sky},
are independent of which finite-time approximation we use.

\subsubsection{Averaging over inclination, polarization, and sky position}
\label{s:averaging}

Using the above expressions for 
$\tilde h_{+,\times}(f;\iota,\phi_0)$ and \eqref{e:R+x}
for $R^{+,\times}(f,\hat k,\psi)$, 
we can calculate the squared response $|\tilde h(f)|^2$ 
averaged over the inclination of the source 
(defined by the inclination and polarization angles
$\iota$ and $\psi$), initial phase $\phi_0$, and sky direction 
$\hat n\equiv -\hat k$.
This is relevant for the case where these quantities 
are not known \emph{a~priori}.
Defining
\begin{multline}
|\tilde h(f;\hat k)|^2
\equiv
\frac{1}{2\pi}\int_0^{2\pi} {\rm d}\phi_0\>
\left(
\frac{1}{4\pi}\int_{0}^{2\pi} {\rm d}\psi\>
\right.
\\
\times
\left.
\int_{-1}^1\>{\rm d}(\cos\iota)\,|\tilde h(f;\hat k,\iota,\psi,\phi_0)|^2
\right)\,,
\end{multline}
it is fairly easy to show that 
\be
\frac{2|\tilde h(f;\hat k)|^2}{T}
\simeq \frac{4}{5}\mathcal{R}(f,\hat k) S_h(f)
\label{e:|htildeI|^2_inc}
\ee
where
\begin{align}
&\mathcal{R}(f,\hat k)\equiv
\frac{1}{2}
\left(|R^+(f,\hat k,0)|^2 + |R^\times(f,\hat k,0)|^2\right)\,,
\label{e:calR_I(f,k)}
\\
&S_h(f) \equiv
\frac{1}{2}\,{h_0^2}
\left[\delta(f-f_0) + \delta(f+f_0)\right]\,.
\label{e:Sh(f)_CW}
\end{align}
The factor of $4/5$ in \eqref{e:|htildeI|^2_inc} 
comes from the average over inclination angles $(\iota,\psi)$;
$\mathcal{R}(f,\hat k)$ encodes the timing residual
response of a pulsar to a plane GW propagating
in direction $\hat k$ averaged over the $(+,\times)$
polarizations and the polarization angle $\psi$; and
$S_h(f)$ is the strain power-spectral
density of a monochromatic GW having frequency $f_0$.
The approximate equality in \eqref{e:|htildeI|^2_inc}
is there because we made the approximation
$\delta_{T}^2(f\mp f_0) \simeq T\delta(f\mp f_0)$
for the product of two finite-time Dirac delta functions.
This allows us to write $S_h(f)$ in terms of ordinary
Dirac delta functions, which are formally singular at $f=\pm f_0$.
But this is not a problem, as $S_h(f)$ will only 
need to be evaluated under an integral sign
for the expected signal-to-noise ratio calculations
that we will perform in Section~\ref{s:matched-filtering}. 
This approximation gives answers that are good to within
$\lesssim 10\%$ for noise power spectral densities that 
don't vary significantly over a frequency bandwidth 
$\Delta f\sim 1/T$ in the neighboorhood of $\pm f_0$.

If we also average over sky location, defining
\begin{multline}
|\tilde h(f)|^2
\equiv
\frac{1}{4\pi}\int {\rm d}^2\Omega_{\hat k}\>
\left(\frac{1}{2\pi}\int_0^{2\pi} {\rm d}\phi_0\>
\right.
\\
\left.
\times\left(
\frac{1}{4\pi}\int_{0}^{2\pi} {\rm d}\psi\>
\int_{-1}^1\>{\rm d}(\cos\iota)\,|\tilde h(f;\hat k,\iota,\psi,\phi_0)|^2
\right)
\right)\,,
\end{multline}
we find
\be
\frac{2|\tilde h(f)|^2}{T}
\simeq \frac{4}{5}\mathcal{R}(f) S_h(f)
\label{e:|htildeI|^2_inc,sky}
\ee
where
\be
\begin{aligned}
\mathcal{R}(f)&\equiv
\frac{1}{8\pi}\int {\rm d}^2\Omega_{\hat k}\>
\left(|R^+(f,\hat k, 0)|^2 + |R^\times(f,\hat k, 0)|^2\right)
\\
&=\frac{1}{12\pi^2 f^2}\,.
\label{e:calR(f)}
\end{aligned}
\ee
Note that the expression for ${\cal R}(f)$ is 
{\em independent} of the direction 
$\hat p$ to the pulsar.
The above expressions will be used later on when defining
the detection sensitivity curves in Section~\ref{s:sensitivity}.

\subsection{Response to a stochastic GW background}
\label{s:stochastic}

For a stochastic GW background, the metric perturbations can be
written as a superposition of plane GWs having
different frequencies $f$, polarizations $\{+,\times\}$,
and propagation directions $\hat k$:
\begin{multline}
h_{ab}(t, \vec x) =
\int {\rm d}^2\Omega_{\hat k}\>
\int_{-\infty}^{\infty}{\rm d}f\>
\left[\tilde h_+(f,\hat k) e^+_{ab}(\hat{k})
\right.
\\
\left.
+\tilde h_\times(f,\hat k) e^\times_{ab}(\hat{k})\right]
e^{i2 \pi f(t - \hat k \cdot \vec x/c)}\,,
\label{e:hab(t,x)_stoch}
\end{multline}
where $e^{+,\times}_{ab}(\hat k)\equiv \epsilon^{+,\times}_{ab}(\hat k,0)$.
This is basically \eqref{e:hab(t,x)_det_freq} but allowing 
for contributions from different propagation direction $\hat k$.
Since we will assume that the sources producing the GW background 
have no preferred polarization direction or symmetry axis, we 
have set  $\psi=0$ and $\iota=0$ in the expansion for $h_{ab}(t,\vec x)$.
The timing residual response of a pulsar to the background
is then
\be
h(t) = \int_{-\infty}^\infty {\rm d}f\>\tilde h(f)e^{i2\pi ft}\,,
\ee
where
\begin{multline}
\tilde h(f) =
\int {\rm d}^2\Omega_{\hat k}\>\left[ 
R^+(f,\hat k,0)\tilde h_+(f;\hat k) 
\right.
\\
\left.
+ R^\times(f,\hat k,0)\tilde h_\times(f;\hat k)\right]
\end{multline}
with $R^{+,\times}(f,\hat k,0)$ given by \eqref{e:R+x}.
As discussed there, we will generally ignore the contribution of 
the pulsar term to the response function, except when calculating 
the auto-correlated power, which will have contributions from
both the Earth-Earth and pulsar-pulsar auto-correlation terms.

The Fourier components $\tilde h_{+,\times}(f;\hat k)$ 
that enter the plane-wave expansion of the metric perturbations
are random fields.
Their quadratic expectation values completely define the 
statistical properties of the background, under the assumption 
that it is Gaussian-distributed.
For simplicity, we will assume that the GW background is 
stationary, unpolarized, and isotropic,%
\footnote{See e.g., \cite{Romano:2016dpx} for a review of analyses that 
drop these assumptions.}
for which $\langle \tilde h_P(f;\hat k)\rangle=0$ and 
\be
\langle \tilde h_P(f;\hat k) \tilde h_{P'}^*(f';\hat k') 
= \frac{1}{16\pi}S_h(f)\delta(f-f')\delta_{PP'}\delta^2(\hat k,\hat k')\,,
\ee
where $P=\{+,\times\}$.
Here $S_h(f)$ is the (one-sided) strain power spectral density of the
background (units of ${\rm strain}{}^2/{\rm Hz}$), which is 
related to the dimensionless energy-density spectrum $\Omega_{\rm gw}(f)$
via
\be
S_h(f) = \frac{3 H_0^2}{2\pi^2} \frac{\Omega_{\rm gw}(f)}{f^3}\,.
\ee
It is also common to describe the background in terms of it 
dimensionless {\em characteristic strain} defined by
\be
h_c(f) \equiv \sqrt{f S_h(f)}= A_{\rm gw}(f/f_{\rm yr})^\alpha\,,
\label{e:hc_power_law}
\ee
where the second equality assumes a power-law form for the background.
Note that for a background produced by the cosmological 
population of SMBH binaries, $\alpha = -2/3$.

\subsubsection{GW contribution to the noise covariance matrix}
\label{s:sgwb_noise}

Using the above expressions for the timing residual response
of a pulsar to a GW background, we can calculate the GW
contribution to the noise covariance matrix when 
cross-correlating timing residuals associated with 
two Earth-pulsar baselines $I$ and $J$.
Denoting the GW contributions to the two sets of timing 
residuals as $h_I(t)$ and $h_J(t)$, respectively,
one can show that the covariance matrix is block-diagonal
with components
\be
C_{h,IJ}
\equiv\langle h_I h_J^T\rangle - 
\langle h_I\rangle \langle h_{J}^T\rangle 
=\chi_{IJ}\,C_h\,,
\label{e:CIJ}
\ee
where
\begin{multline}
\chi_{IJ} \equiv
\frac{1}{2} +
\frac{3}{2}\left(\frac{1-\hat{p}_I \cdot \hat{p}_J}{2}\right)
\left[\ln\left(\frac{1-\hat{p}_I \cdot \hat{p}_J}{2}\right) - \frac{1}{6}\right]
\\
+\frac{1}{2}\,\delta_{IJ}\,,
\label{e:HD}
\end{multline}
and
\begin{align}
&C_{h,ij} = \int_0^{f_{\rm Nyq}}{\rm d}f\> \cos[2\pi f(t_i-t_i)]P_h(f)\,,
\label{e:Ch}
\\
&P_h(f) = {\cal R}(f) S_h(f)
=\frac{A_{\rm gw}^2}{12\pi ^2}\left(\frac{f}{f_{\rm yr}}\right)^{2\alpha}f^{-3}\,.
\label{e:Ph}
\end{align}
The full noise covariance matrix, which includes contributions
instrinsic to the pulsar and to the measurement process, is also
block-diagonal with components 
\be
C_{IJ} = \delta_{IJ} C_{n,I} + C_{h,IJ}\,.
\label{e:C_IJ}
\ee
Here $C_{n,I}$ is given by \eqref{e:Ch}, but 
with the pulsar noise power spectral density $P_{n_I}(f)$ replacing $P_h(f)$.
This last equation assumes that the noise contributions 
associated with different pulsars are not correlated with one 
another.

\begin{figure}[h!tbp]
\centering
\includegraphics[width=\columnwidth]{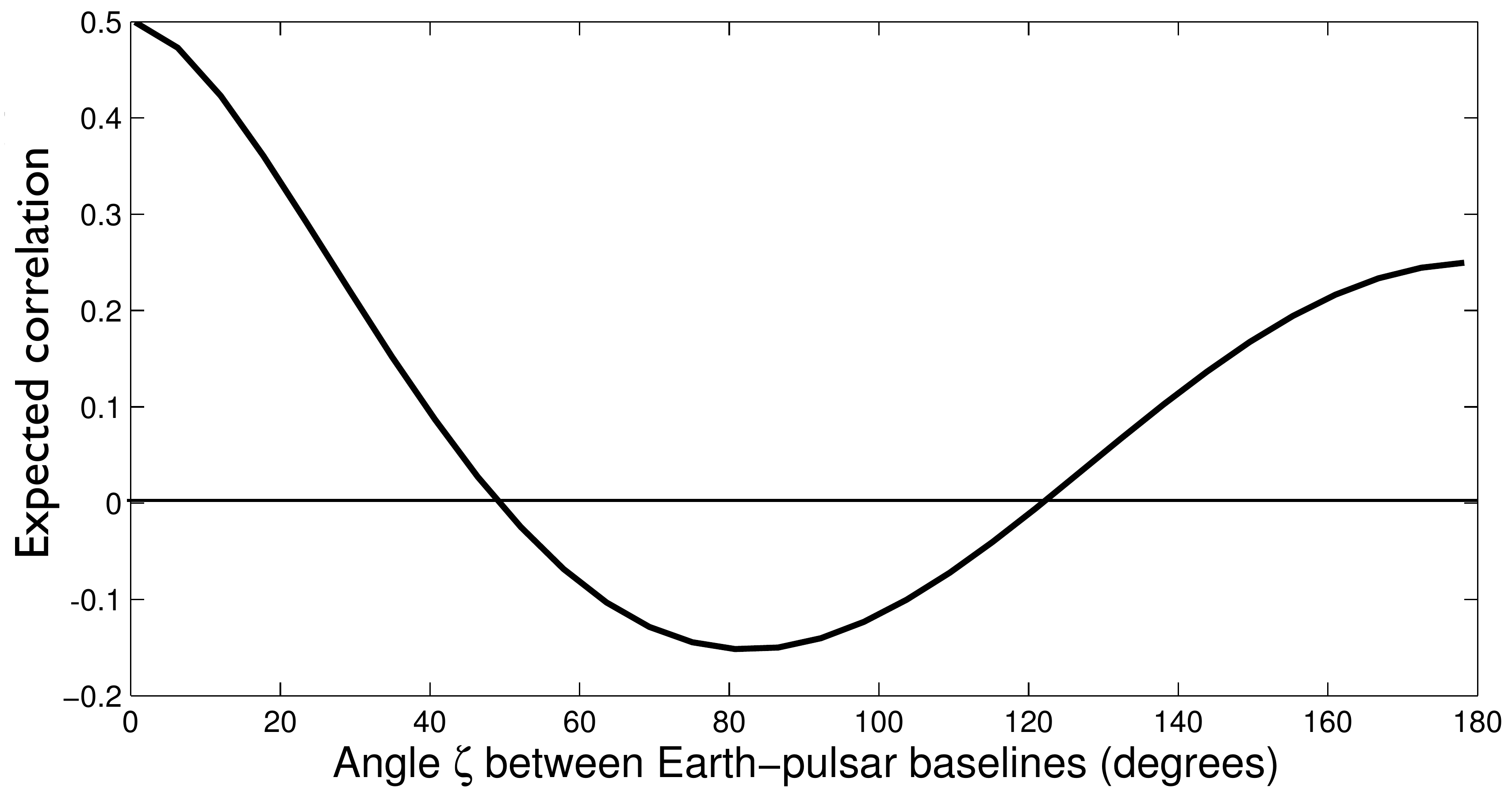}
\caption{Hellings and Downs curve.
Plotted is the expected correlation for the timing residuals 
induced in a pair of distinct Earth-pulsar baselines by an 
unpolarized, isotropic GW background.}
\label{f:HD}
\end{figure}

The quantity $\chi_{IJ}\equiv \chi(\zeta_{IJ})$ defined in 
\eqref{e:HD} is the Hellings and Downs factor~\cite{Hellings:1983fr}
for a pair of pulsars separated by angle 
$\zeta_{IJ}=\cos^{-1}(\hat p_I\cdot\hat p_J)$ 
(see Figure~\ref{f:HD}).
It arises when cross-correlating the GW-induced timing
residuals for an unpolarized, isotropic GW background.
Note that $\chi_{IJ}$ has been normalized such that $\chi_{II}=1$ 
(for a single pulsar).
%

%% file: manuscript/sensitivity.tex
\section{Sensitivity curves}
\label{s:sensitivity}

Ultimately, a detection sensitivity curve should tell us
how likely it is to detect a particular type of GW signal.
So it should depend not only on the properties of the noise
in the detector, but also on the type of signal that one
is searching for and the method that one uses to search for it.
So here we extend the formalism of the previous two sections
to define sensitivity curves for searches for 
a deterministic GW signal from a circular binary and
an unpolarized, isotropic stochastic GW background.
We begin by writing down expressions for the optimal detection
statistics for these two different sources and their
corresponding expected signal-to-noise ratios (SNRs).
We will see that from these expected SNRs, we can read off
an {\em effective} strain-noise power spectral density, which has
the interpretation of a detection sensitivity curve.

\subsection{Matched filtering for a deterministic GW signal}
\label{s:matched-filtering}

For a deterministic GW signal,
we can use the method of matched
filtering to construct an optimal detection statistic. 
This method has been used extensively in the PTA literature, 
\cite{bps+2016,bs12,Ellis:2012zv,yhj+10} and is also the 
basis for the approximate deterministic sensitivity 
curves in \cite{Moore:2014eua}.
Letting $Q_I$ denote the filter function for pulsar
$I$ (where $I=1,2,\cdots, N_{\rm p}$), we define
\be
\hat{\mathcal{S}}
\equiv \sum_I Q_I^T r_I = \sum_I\sum_{\alpha} Q_{I\alpha} r_{I\alpha}\,,
\ee
where $r_I \equiv G_I^T\delta t_I$ are the TMM residuals
for pulsar $I$.
The filter function is determined by maximizing
the expected signal-to-noise ratio,
$\rho\equiv\mu/\sigma$, of $\hat{\mathcal{S}}$.
The expectation value of $\hat{\mathcal{S}}$ is
given by
\be
\mu \equiv \langle \hat{\mathcal{S}} \rangle
= \sum_I Q_I^T G_I^T h_I(\theta)
\ee
and its variance is given by
\be
\sigma^2 \equiv
\langle  \hat{\mathcal{S}} ^2 \rangle -
\langle  \hat{\mathcal{S}}\rangle ^2
=
\sum_I
Q_I^T \Sigma_{n,I} Q_I\,,
\ee
where
$\Sigma_{n, I}\equiv G_I^T C_{n,I} G_I$
is the noise covariance matrix for $r_I$.
This result for the variance assumes that the only GW
contribution to the timing residuals is from a deterministic
GW source, and not from a stochastic GW background.
The presence of a stochastic background would
contribute to both the diagonal and off-diagonal
block matrices (see \eqref{e:C_IJ}).
In what follows, we will assume that the off-diagonal terms
are small compared to the diagonal (auto-correlated) terms.
But we will replace $\Sigma_{n,I}$
by $\Sigma_I\equiv G_I^T C_I G_I$, where
$C_I\equiv C_{n,I}+C_h$,
thereby allowing a stochastic background to contribute
to the auto-correlated  noise (sometimes called GW
{\em self-noise}).

Using the above results for the mean and variance
of $\hat{\mathcal{S}}$,
the square of the expected signal-to-noise ratio is
\be
\rho^2 \equiv
\frac{\mu^2}{\sigma^2} =
\frac{\sum_{I,J}
Q^T_{I}\,G^T_I h_I(\theta)
Q^T_J\, G^T_J h_J(\theta)}
{\sum_K Q_K^T \Sigma_K Q_{K}}\,,
\ee
with the optimal filter given by
\be
\frac{\delta\rho^2}{\delta Q_I} = 0
\quad\Rightarrow\quad
Q_{I} =
\Sigma_{I}^{-1}
G^T_I h_I(\theta)\,.
\ee
Note that $Q_I$  is a noise-weighted version of the
TMM signal waveform, as expected for a matched-filter
statistic.
Using this expression, the expected signal-to-noise ratio
becomes
\be
\begin{aligned}
\rho^2(\theta)
&= \sum_I h_I(\theta)^T G_I
\Sigma_{I}^{-1} G_I^T h_I(\theta)
\\
&= \sum_I h_I(\theta)^T G_I
(G^T_I C_I G_I)^{-1} G_I^T h_I(\theta)\,.
\end{aligned}
\ee
This last expression can be evaluated in the
frequency domain by using \eqref{e:Ninvff'} for
${\cal N}_I^{-1}(f,f')$ and restricting to the diagonal
component ${\cal N}^{-1}(f)$ as discussed in
Section~\ref{s:Ninvff'}:
\be
\rho^2(\theta)\simeq
4\int_0^{f_{\rm Nyq}}{\rm d}f\>\sum_I
|\tilde h_I(f;\theta)|^2\,{\cal N}_I^{-1}(f)\,.
\ee
Recall that $\theta$ denote the set of GW
parameters.
For the case of a circular binary discussed
in Section~\ref{s:circular_binaries},
$\theta=\{\hat k,\iota, \psi, \phi_0\}$.

\subsubsection{Detection sensitivity curve for sky and
inclination-averaged sources}
\label{s:detection_angle_averaging}

\begin{figure}[h!tbp]
\centering
\includegraphics[width=\columnwidth]{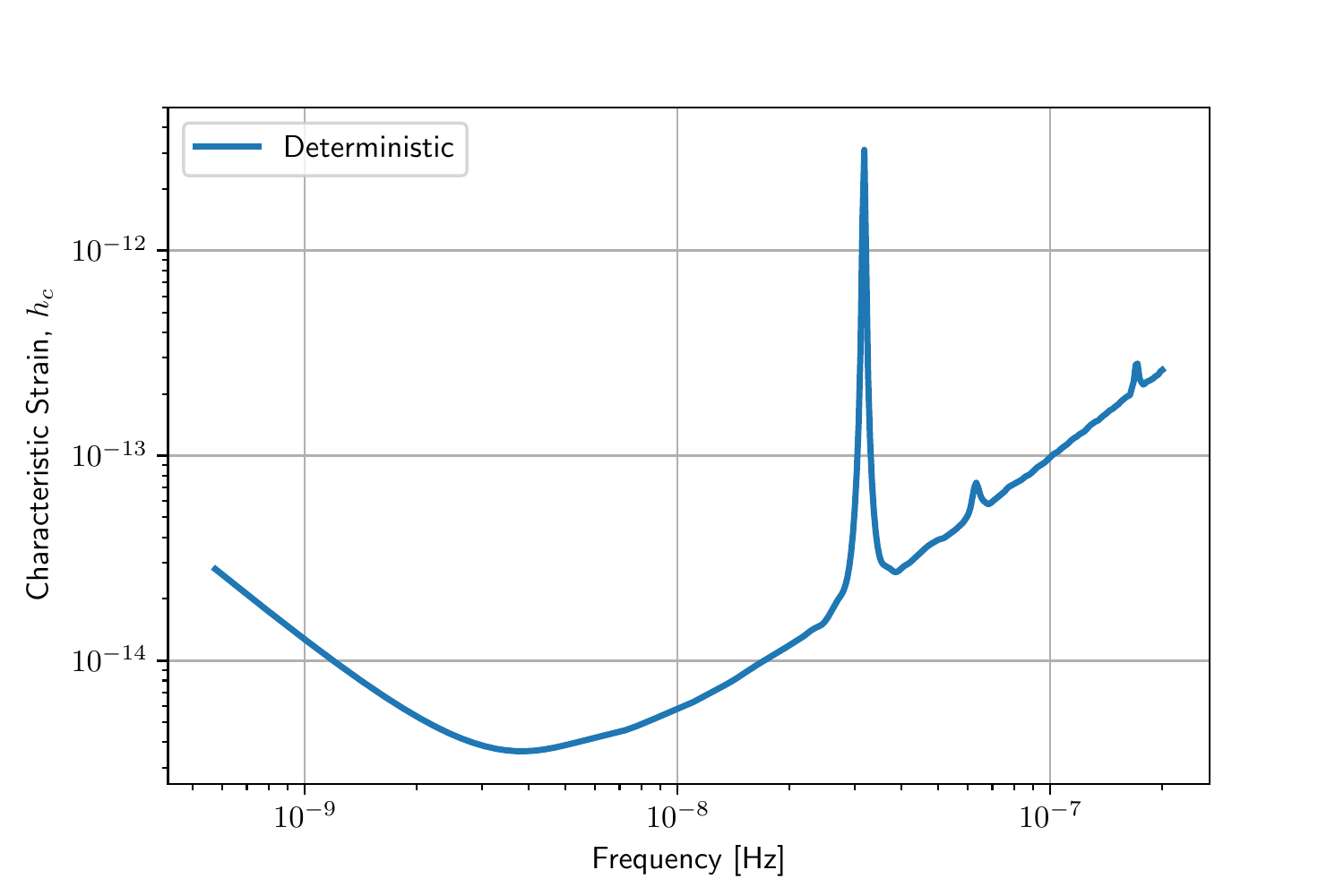}
\caption{Sensitivity curve for a single deterministic
GW source averaged over its initial phase, inclination,
and sky location.
This plot was constructed using the NANOGrav 11-year data.}
\label{f:Seff_inc_sky}
\end{figure}

To proceed further, we first consider the case of GWs
from a single binary system averaged over the initial phase,
inclination of the source, as well as its sky location.
Using \eqref{e:|htildeI|^2_inc,sky} for $|\tilde h_I(f)|^2$, 
we have
\be
\begin{aligned}
\langle\rho^2\rangle_{\rm inc,\, sky}
&\simeq 4\int_0^{f_{\rm Nyq}} {\rm d}f\>\sum_I
\frac{T_I}{2}\frac{4}{5}
\mathcal{R}(f) S_h(f)\,
\mathcal{N}_{I}^{-1}(f)
\\
&= 2 T_{\rm obs}\int_0^{f_{\rm Nyq}} {\rm d}f\>\frac{S_h(f)}{S_{\rm eff}(f)}\,,
\end{aligned}
\ee
where
\begin{align}
&{S_{\rm eff}(f)}
\equiv \left(
\frac{4}{5}\sum_I \frac{T_I}{T_{\rm obs}}\frac{1}{S_I(f)}\right)^{-1}\,,
\label{e:Seff_inc,sky}
\\
&S_I(f) 
\equiv \frac{1}{\mathcal{N}^{-1}_I(f)\mathcal{R}(f)}\,.
\label{e:SI(f)}
\end{align}
Here, $S_I(f)$ is the strain-noise power spectral
density for pulsar $I$, and $S_{\rm eff}(f)$
is an {\em effective} strain noise power spectral
density for an array of pulsars.
Given how $S_{\rm eff}(f)$ appears in the expression for the
expected signal-to-noise ratio, we will use it, or
its dimensionless characteristic strain,
\be
h_{\rm eff}(f) \equiv \sqrt{f S_{\rm eff}(f)}\,
\ee
as a {\em sensitivity curve} for detecting a deterministic
GW source averaged over its initial phase, inclination, and
sky location.
A plot of $S_{\rm eff}(f)$ for the array of pulsars in 
the NANOGrav 11-year data \cite{Arzoumanian:2018saf}
is shown in Figure~\ref{f:Seff_inc_sky}.
Note that for a monochromatic source, $S_h(f)$ has a very
simple form given by \eqref{e:Sh(f)_CW}, which implies
\be
\bar\rho
\equiv \sqrt{\langle \rho^2\rangle_{\rm inc,\,sky}}
\simeq h_0\sqrt{\frac{T_{\rm obs}}{S_{\rm eff}(f_0)}}\,.
\label{e:bar_rho}
\ee
%

\subsubsection{SNR and characteristic amplitude sky maps for
inclination-averaged sources}
\label{s:skymaps_angle_averaging}

\begin{figure}[h!tbp]
\centering
\includegraphics[width=\columnwidth]{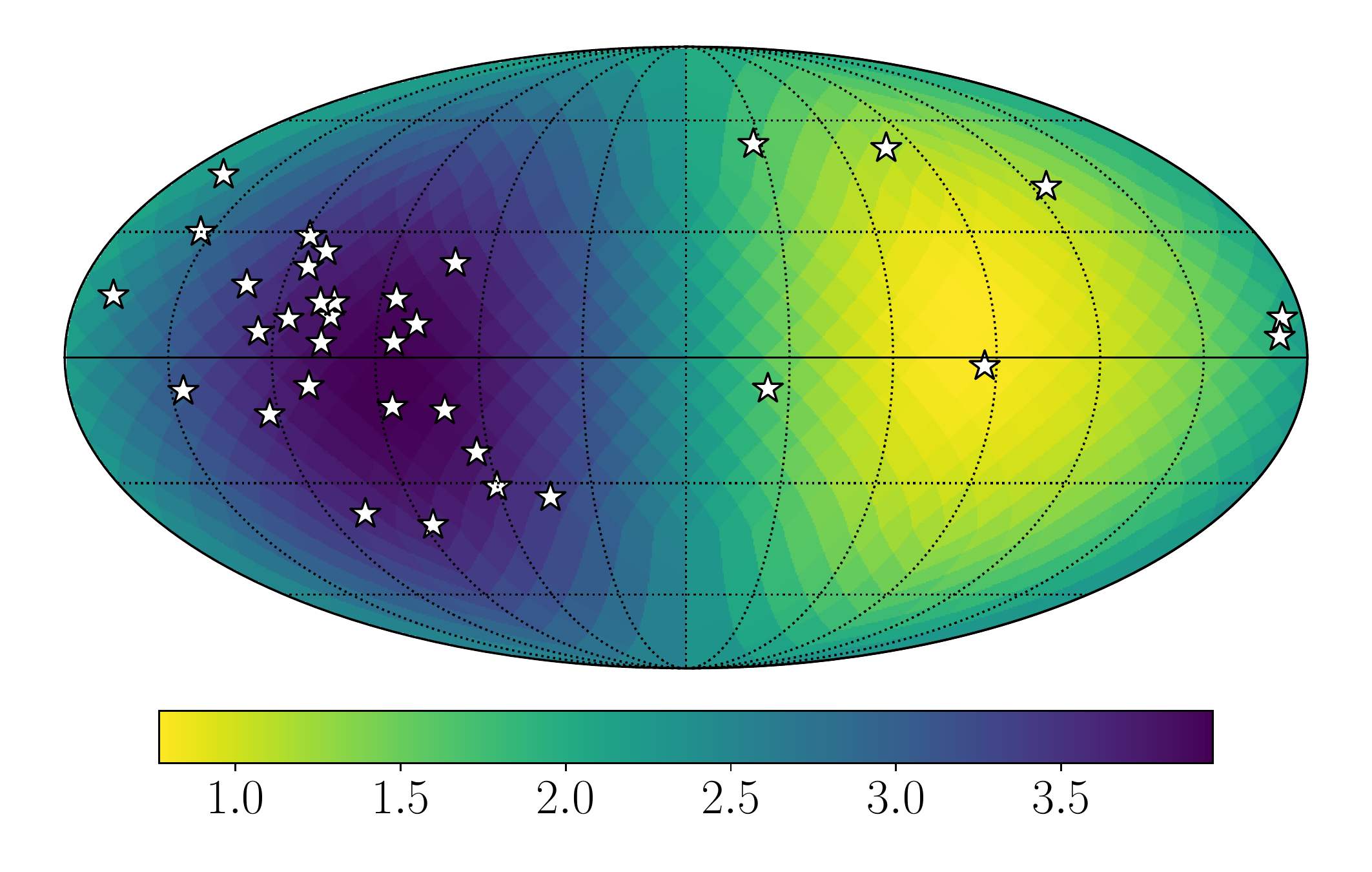}
\caption{Sky map of the expected matched-filter signal-to-noise
ratio $\rho(\hat n)$ for a monochromatic circular binary
(GW frequency $f_0=8~{\rm nHz}$) consisting of a pair of
$10^9$ solar-mass BHs at a luminosity distance of 100~Mpc.
This plot was constructed using the NANOGrav 11-year data.}
\label{f:snr_map}
\end{figure}

If we average over initial phase and source inclination,
but not over sky location, cf.~\eqref{e:|htildeI|^2_inc}
for $|\tilde h_I(f,\hat k)|^2$, we obtain
\be
\begin{aligned}
\langle\rho^2(\hat k)\rangle_{\rm inc}
&\simeq 4\int_0^{f_{\rm Nyq}} {\rm d}f\>\sum_I
\frac{T_I}{2}\frac{4}{5}
\mathcal{R}_I(f,\hat k)S_h(f)\,
\mathcal{N}_{I}^{-1}(f)
\\
&= 2 T_{\rm obs}\int_0^{f_{\rm Nyq}} {\rm d}f\>
\frac{S_h(f)}{S_{\rm eff}(f,\hat k)}\,,
\label{e:rho2(k)}
\end{aligned}
\ee
where
\begin{align}
&S_{\rm eff}(f,\hat k)\equiv \left(
\frac{4}{5}\sum_I \frac{T_I}{T_{\rm obs}}\frac{1}{S_I(f,\hat k)}\right)^{-1}\,,
\label{e:Seff_inc}
\\
&S_I(f,\hat k) \equiv \frac{1}{\mathcal{N}^{-1}_I(f)\mathcal{R}_I(f,\hat k)}\,,
\label{e:SI(f,k)}
\end{align}
with $\mathcal{R}_I(f,\hat k)$ given by \eqref{e:calR_I(f,k)}.
These expressions are analogous to \eqref{e:Seff_inc,sky},
but with added dependence on the propagation direction $\hat k$ of the GW.
It turns out that we can factor out the $\hat k$ dependence
on the right-hand side of the above expression for $S_{\rm eff}(f,\hat k)$
if we ignore the frequency-dependent part of the pulsar-term
contribution to $|R_I^P(f,\hat k, 0)|^2$, as discussed in the
context of \eqref{e:pulsar-pulsar}.
Making this approximation,
\begin{multline}
S_{\rm eff}(f,\hat k)
\simeq
\left(\frac{12}{5}\sum_I \frac{T_I}{T_{\rm obs}}
\frac{1}{S_I(f)}\,
\right.
\\
\left.
\times
\left[(F^+_I(\hat k))^2 + (F^\times_I(\hat k))^2\right]
\right)^{-1}\,,
\label{e:Seff(f,k)}
\end{multline}
where $F_I^{+,\times}(\hat k)$ are defined by
\be
F_I^{+,\times}(\hat k) \equiv
\frac{1}{2}\frac{\hat p_I^a\hat p_I^b}{1+\hat p_I\cdot\hat k}\,
e_{ab}^{+,\times}(\hat k)\,.
\ee
As before, it is easy to do the integral over frequency
for a monochromatic source, for which $S_h(f)$ is given
by \eqref{e:Sh(f)_CW}.
The result is
\be
\rho(\hat n)\equiv
\sqrt{\langle \rho^2(\hat k)\rangle_{\rm inc}}
\simeq h_0 \sqrt{\frac{T_{\rm obs}}{S_{\rm eff}(f_0,\hat k)}}\,,
\label{e:rho(n)}
\ee
where the direction $\hat n$ of the source on the sky
is opposite the direction of GW propagation, $\hat n=-\hat k$.
A plot of $\rho(\hat n)$ for a pair of $10^9$~solar-mass
BHs at a luminosity distance of 100~Mpc,
emitting monochromatic GWs at the frequency $f_0 = 8~{\rm nHz}$
is shown in Figure~\ref{f:snr_map}.

\begin{figure}[h!tbp]
\centering
\subfigure[]{\includegraphics[width=\columnwidth]
{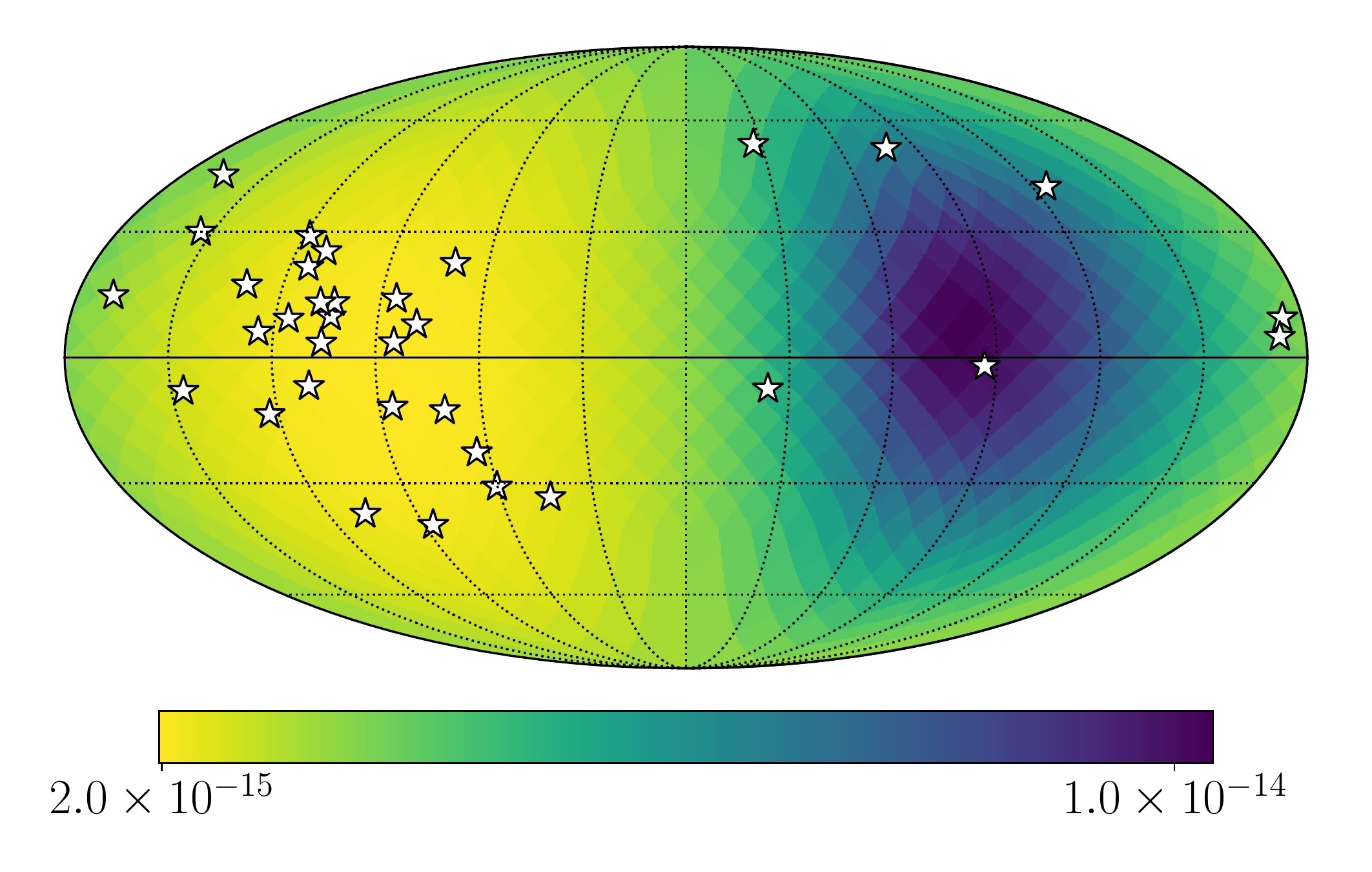}}
\subfigure[]{\includegraphics[width=\columnwidth]
{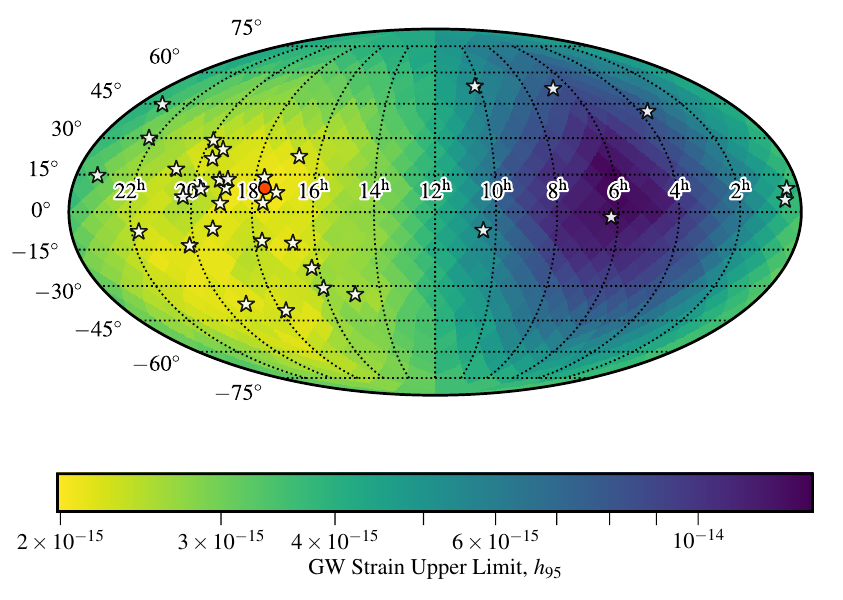}}
\caption{Panel (a): Sensitivity sky map for the strain
amplitude of a monochromatic continuous-wave source,
calculated using the NANOGrav 11-year data~\cite{Arzoumanian:2018saf}.
For this plot, we have taken $f_0=8~{\rm nHz}$ and $\rho=2$.
Panel (b): For comparison, a 95\% confidence-level upper
limit sky map taken from the
NANOGrav 11-year single-source paper~\cite{Aggarwal:2018mgp}.}
\label{f:h0_map}
\end{figure}

Finally, it is a simple matter to recast the form of the
sky map so that we solve \eqref{e:rho(n)}
for the strain amplitude $h_0$ of a monochromatic binary,
cf.~\eqref{e:h0}, that would
produce a particular value of the signal-to-noise ratio
$\rho$:
\be
h_0(\hat n)
= {\rho}\sqrt{\frac{S_{\rm eff}(f_0,\hat k)}{T_{\rm obs}}}\,.
\ee
A sky map of $h_0(\hat n)$ is shown in panel (a) of
Figure~\ref{f:h0_map} for $\rho=2$ using the
NANOGrav 11-year data.
For comparison, panel (b) shows the actual 
95\% confidence-level upper
limit map taken from the NANOGrav 11-year
single-source paper~\cite{Aggarwal:2018mgp}.
%

\subsection{Single-pulsar characteristic strain noise curves}
\label{s:single_pulsar}

\begin{figure}[h!tbp]
\centering
\includegraphics[width=0.95\columnwidth]{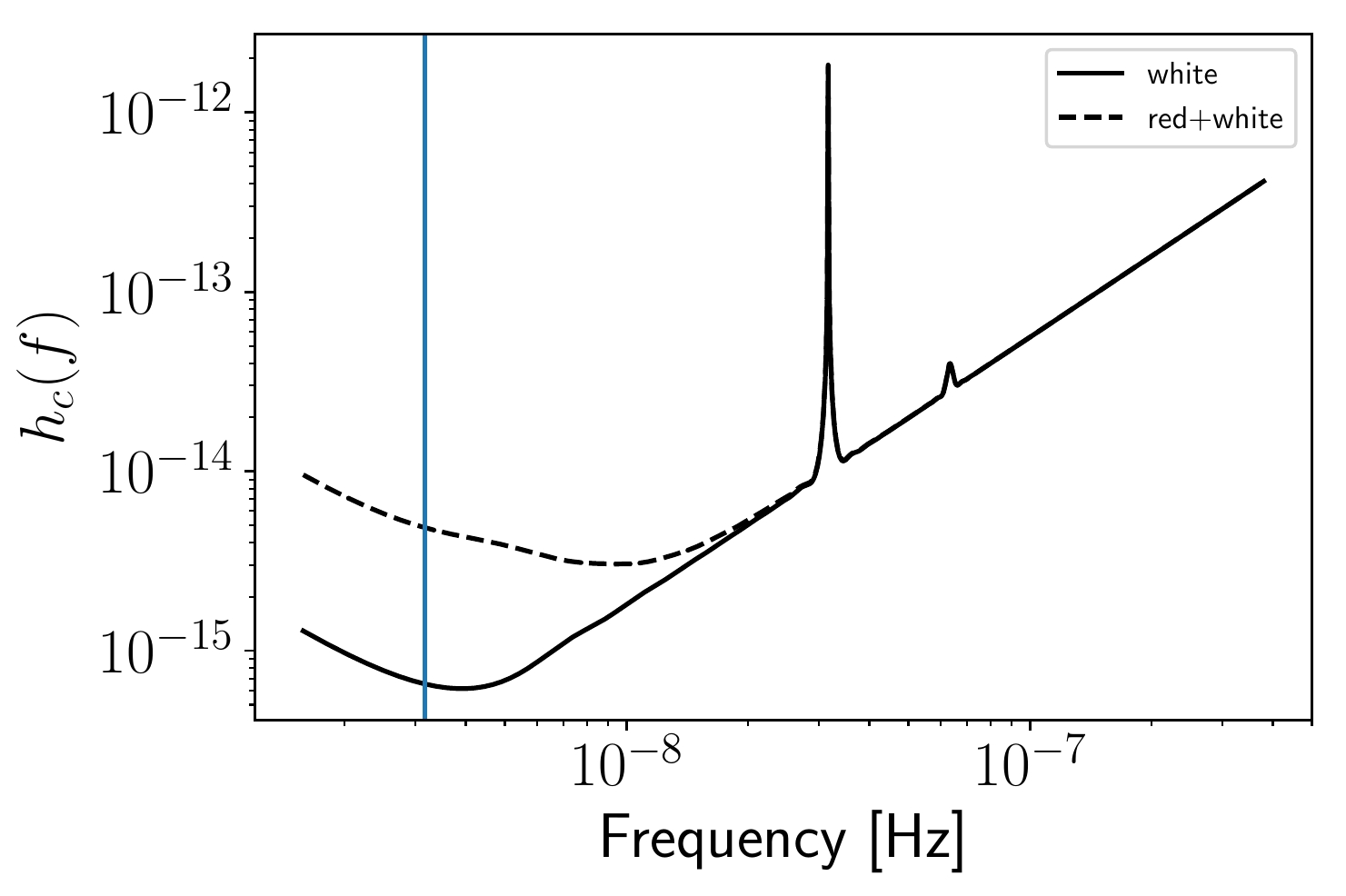}
\caption{Single-pulsar characteristic strain-noise sensitivity 
curves for the
simple quadratic spin-down timing model fit described in the main
text and for white noise (solid curve) and red+white noise
(dashed curve).  The vertical blue line corresponds to a frequency
of $1/T$.}
\label{f:hc_comparison}
\end{figure}

\begin{figure*}
\centering
\includegraphics[width=\textwidth]{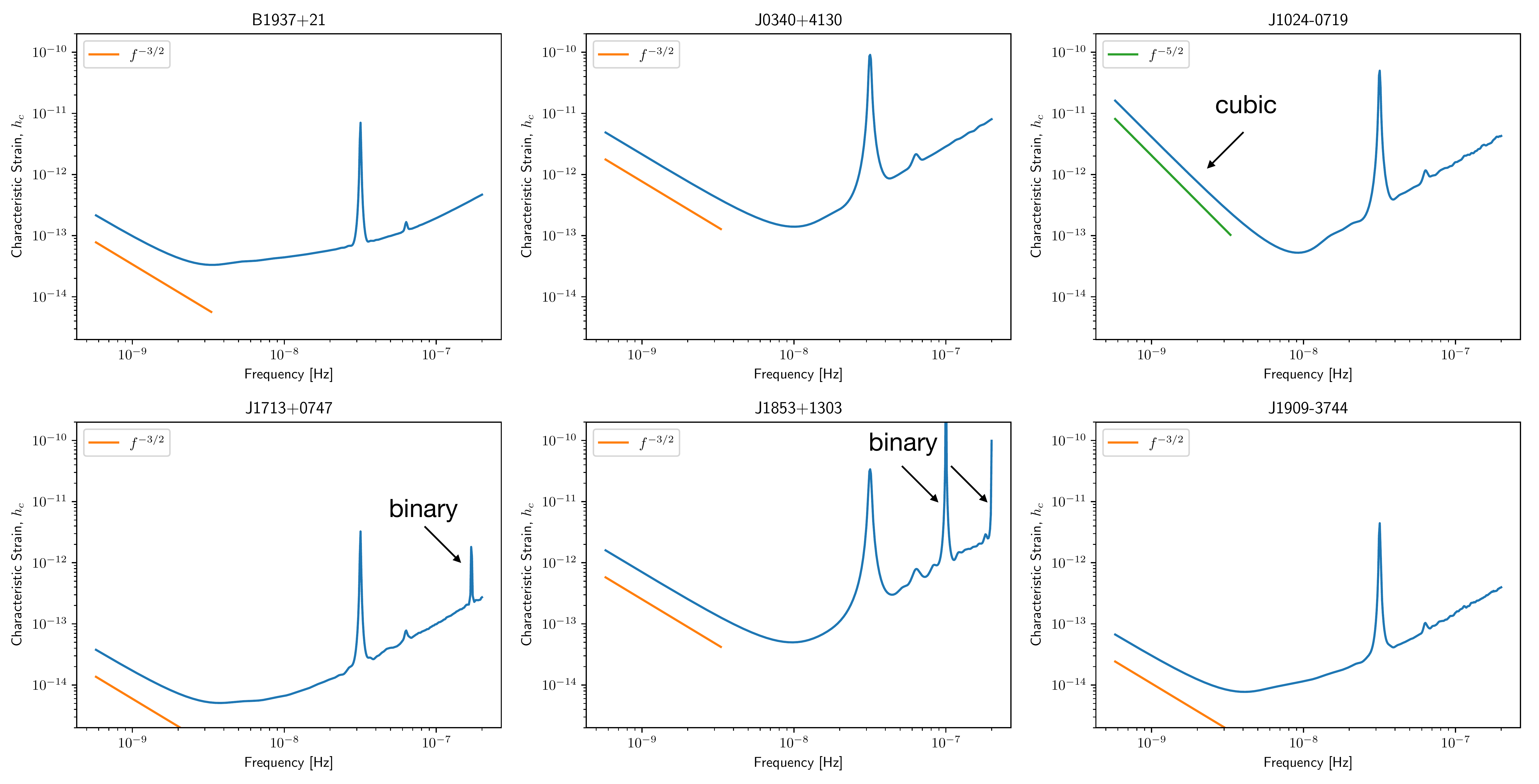}
\caption{Single-pulsar characteristic strain-noise sensitivity 
curves for a subset of NANOGrav 11-year pulsars.
The $\sim\!f^{-5/2}$ behavior for PSR J1024-0719 is evidence of a fit to
a cubic spin-down model for the pulsar spin frequency. The cubic term in the fit is needed due to an acceleration of the pulsar, evident in the TOAs from its unusually-long binary period \cite{Arzoumanian:2018saf, Kaplan:2016ymq}.
The additional spikes seen for J1713+0747 and J1853+1303 show
that the pulsar is in a binary system; the second binary spike
for J1853+1303 is the second harmonic of the binary orbital
frequency.}
\label{f:ng11yr_single_pulsar}
\end{figure*}

For an individual pulsar, we will use the characteristic
strain
\be
h_{c,I}(f) \equiv \sqrt{f S_I(f)}\,,
\quad
S_I(f)\equiv \frac{1}{{\cal N}_I^{-1}(f){\cal R}(f)}\,,
\ee
to characterize its polarization and sky-averaged 
sensitivity; see~\eqref{e:SI(f)}.
Plots of single-pulsar characteristic strain-noise
sensitivity curves for the simple quadratic spin-down model described
in Section~\ref{s:TOAs_residuals} and for both white and red+white
noise are shown in Figure~\ref{f:hc_comparison}.
More realistic single-pulsar strain-noise sensitivity curves can 
be constructed using a subset of the NANOGrav 11-year pulsars
(Figure~\ref{f:ng11yr_single_pulsar}) \cite{Arzoumanian:2018saf}.
These pulsars have noise
contributions specified by the parameters EQUAD, ECORR, and
EFAC~\cite{Arzoumanian:2018saf,Lam:2016iie,Arzoumanian:2014gja}, which are denoted by
$Q$, $J_{ij}$, and $F$ in the
following expression for the noise covariance matrix:
\be
C_{n,ij} = F^2 \,[\sigma_i^2\,\delta_{ij} + Q^2\delta_{ij}] + J_{ij}\,.
\ee
Here $\sigma_i^2$ are individual TOA errors, which
are associated with the finite-signal-to-noise ratio
determination of the pulse arrival times (obtained by
correlating the observed pulses with a pulse template).
EQUAD  are white noise
contributions to the covariance matrix that add in
quadrature with the TOA errors. EFAC is an
overall scale factor that can be used to adjust the
overall uncertainty if necessary. ECORR are noise contributions
that are correlated within an observing epoch, but not from epoch to
epoch. Hence the ECORR contributions to the covariance matrix are
block diagonal. Red noise, modeled as a power law, was added for those
pulsars that show significant detections in the NANOGrav 11-year data
set \cite{Arzoumanian:2018saf}. In
Figure~\ref{f:ng11yr_single_pulsar}, B1937+21, J1713+0747 and
J1909-3744 have injections of red noise. This can be distinguished by
the ``flatter'' appearance of the sensitivity curves around the
minimum, as compared to the other pulsars. For a detailed list of
noise parameters, and to see which pulsars have significant detections
of red noise, consult Table~2 in \cite{Arzoumanian:2018saf}.

\begin{figure}[h!tbp]
\centering
\includegraphics[width=0.9\columnwidth]{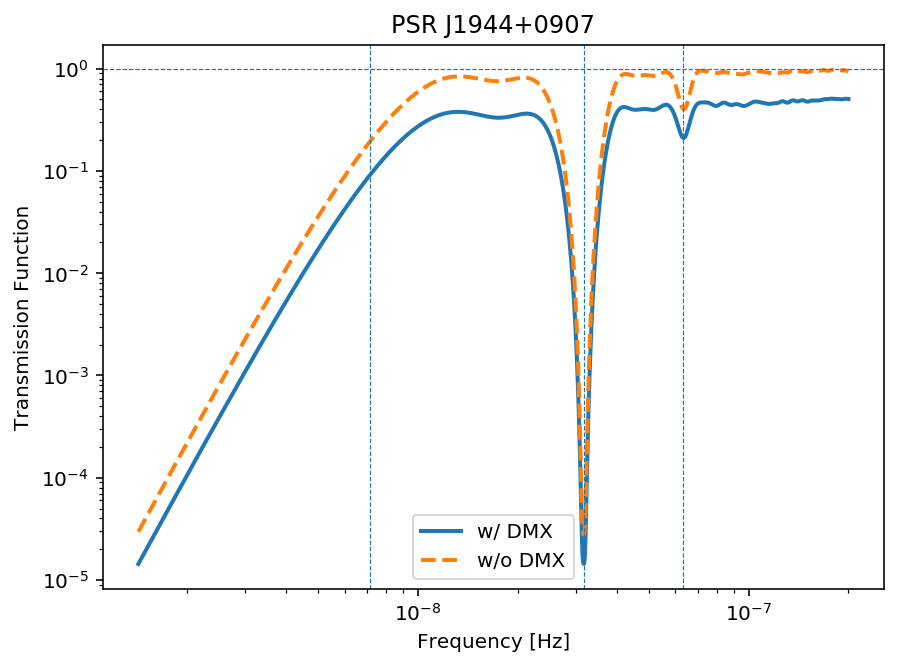}
\caption{Plots of transmission functions showing the effect of
including time-dependent dispersion measure (DMX) into the
timing model fit.
Including DMX in the timing model leads to broadband absorption of power
(solid blue curve) relative to that for a timing model without DMX.}
\label{f:transmission_DMX}
\end{figure}

The NANOGrav 11-year pulsars also have more complicated timing
model fits than the simple quadratic spin-down model
described in Section~\ref{s:TOAs_residuals}.
In Figure~\ref{f:ng11yr_single_pulsar}, one can
see that pulsar J1024-0719 is fit to a cubic
spin-down model, leading to a steeper
frequency-dependence ($\sim\!f^{-5/2}$) at low frequencies.
One also sees that J1713+0747 and J1853+1303 are in
binary systems: there are additional spikes at the
binary orbital frequency and twice the binary orbital
frequency for J1853+1303.
Finally, these pulsars have timing models that also
include fits to a piecewise, time-dependent dispersion measure fluctuation
(DMX), which is associated with perturbations of the dispersion of
the radio pulses as they propagate through the interstellar
medium from the pulsar to a radio receiver on Earth.
(The lower-frequency components of a pulse are
delayed more than the higher-frequency components.)
Fitting to DMX in the timing model leads to
broadband absorption of power relative to a timing
model that doesn't fit for DMX.
Figure~\ref{f:transmission_DMX} shows plots of the
transmission function for NANOGrav pulsar J1944+0907,
with and without DMX included in the timing model.

\subsection{Optimal cross-correlation statistic for a
stochastic GW background}
\label{s:optimal_GWB}

The derivation of the optimal cross-correlation statistic
for a stochastic GW background is similar to that
presented above for a single deterministic GW, expect 
that we work with data from pairs of pulsars.
Starting with a single distinct pair, labeled by $I$
and $J$, we define
\be
\hat{\mathcal{S}}_{IJ} \equiv
r_I^T Q\, r_J\,,
\ee
where $r_I$ and $r_J$ are the TMM residuals
for pulsars $I$ and $J$
(assuming that we have already fit for all deterministic
GW sources), and $Q$ is an $m_I\times m_J$ matrix,
where
$m_I\equiv N_I-N_{{\rm par},I}$, etc.
As before, we determine the filter function $Q$
by maximizing the signal-to-noise ratio of $\hat{\mathcal{S}}_{IJ}$.
Similar derivations appear in the
literature~\cite{Allen-Romano:1999, Anholm:2008wy, Ellis:2012zv, Chamberlin:2014ria, viet17}.
The final result for the optimal filter is
\be
Q \propto \,\Sigma_I^{-1}
\Sigma_{IJ}\Sigma_J^{-1}\,,
\ee
where
\be
\begin{aligned}
&\Sigma_I
\equiv G_I^T(C_{n,I} + C_h)G_I\,,
\\
&\Sigma_{IJ}
\equiv \chi_{IJ}\,G_I^T C_hG_J\,.
\end{aligned}
\ee
The expected squared signal-to-noise ratio for this
optimal choice of $Q$ is then
\begin{equation}
\rho_{IJ}^2 =
{\rm Tr}\left[\Sigma_{JI}\,\Sigma_I^{-1}
\Sigma_{IJ}\Sigma_J^{-1}\right]\,.
\end{equation}
The above calculation assumes that we are in the
weak-signal limit where the cross-correlation terms
are assumed to be negligible compared to auto-correlation
terms (i.e., we assume that the GW signal power is much less than
that for the intrinsic pulsar and measurement noise).

We can then combine the signal-to-noise ratios for
each distinct pair in quadrature since,
in the weak-signal limit,
there is negligible correlation between these estimators:
\begin{equation}
\rho^2 \simeq
\sum_{I}\sum_{J>I} \rho_{IJ}^2.
\end{equation}
As we saw for deterministic GWs, it is useful to write
the above expression for the expected squared signal-to-noise
ratio in the frequency domain.
Proceeding as we did there, we find
\begin{multline}
\rho^2
\simeq \sum_I \sum_{J>I}
2 T_{IJ}\chi_{IJ}^2
\\
\times
\int_{0}^{f_{\rm Nyq}}{\rm d}f\>
S^2_h(f) \mathcal{R}^2(f)
\mathcal{N}^{-1}_I(f)
\mathcal{N}^{-1}_J(f)\,,
\end{multline}
where $P_h(f) = \mathcal{R}(f)S_h(f)$, and
where $\mathcal{N}_I^{-1}(f)$ is defined by
\eqref{e:Ninvf}.
This suggests defining the
following {\em effective} strain-noise power
spectral density for the whole PTA:
\be
S_{\rm eff}(f)
=
\left(\sum_I\sum_{J>I}
\frac{T_{IJ}}{T_{\rm obs}}
\frac{\chi^2_{IJ}}{S_I(f)S_J(f)}
\right)^{-1/2}\,,
\label{e:Seff_gwb}
\ee
which includes contributions from the Hellings and
Downs factors $\chi_{IJ}$ and the individual pulsar
strain-noise power spectral densities
$S_I(f)\equiv 1/(\mathcal{N}^{-1}_I(f)\mathcal{R}(f))$.
Note that $S_{\rm eff}(f)$ has dimensions of strain${}^2$/Hz,
and that
\be
\rho^2\simeq
2T_{\rm obs}\int_0^{f_{\rm Nyq}}{\rm d}f\>
\frac{S_h^2(f)}{S_{\rm eff}^2(f)}
\label{e:rho2_gwb}
\ee
in terms of $S_{\rm eff}(f)$.

\begin{figure}[h!tbp]
\centering
\includegraphics[width=\columnwidth]{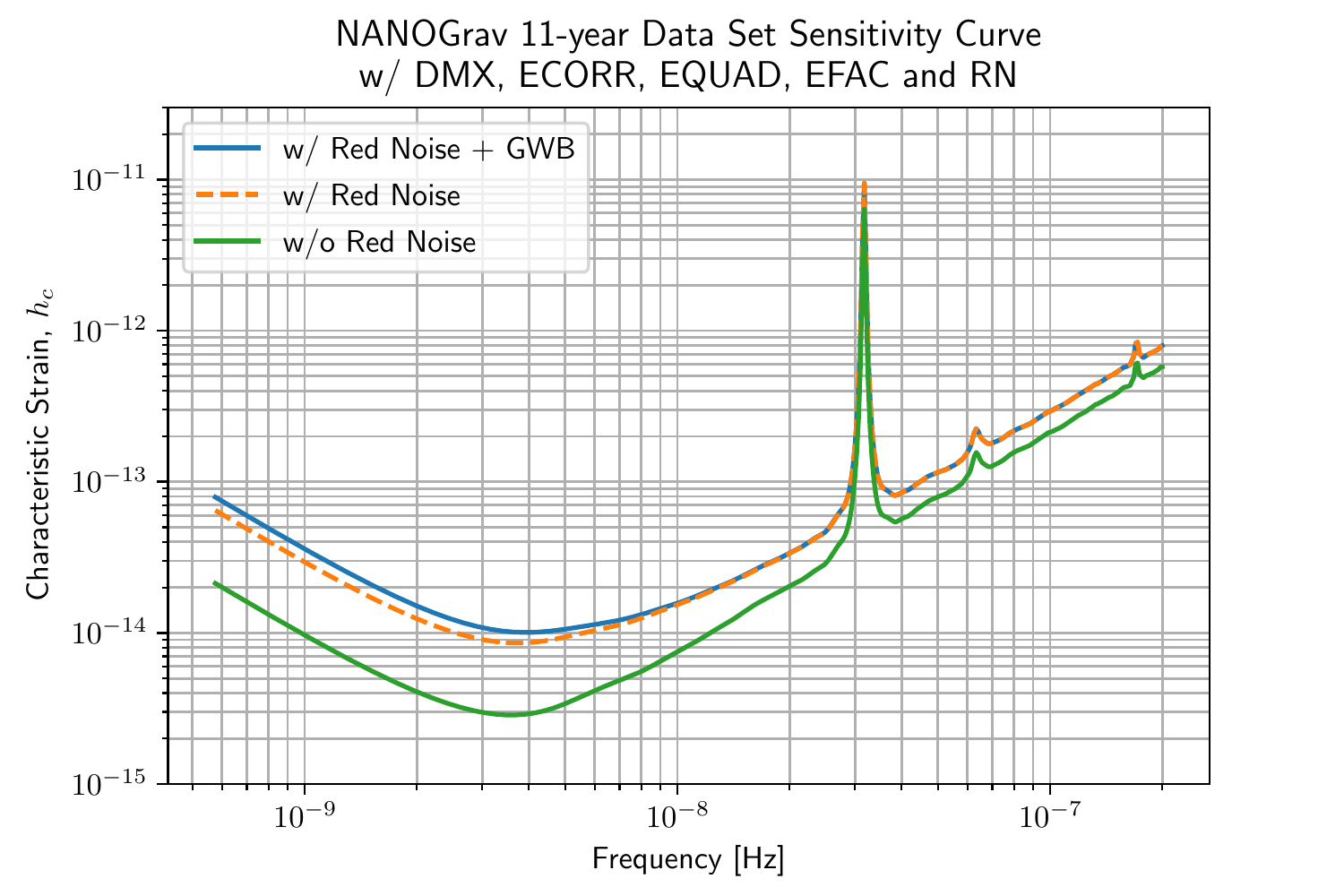}
\caption{Comparison of stochastic sensitivity curves (effective
characteristic strain noise) for the NANOGrav 11-year PTA.
All the curves include realistic pulsar noise characteristics and
individual timing model fits.
The blue curve includes a contribution to the auto-power spectra,
produced by a GWB at the level of $A_{\rm gwb}=1\times 10^{-16}$.
The dashed-orange curve shows the sensitivity without including
the GWB, and the green curve shows what happens if you also ignore
the red noise contributions to the noise covariance matrices.}
\label{f:ng11yr_gwb}
\end{figure}
A plot comparing dimensionless charateristic strain curves
$h_c(f)\equiv \sqrt{f S_{\rm eff}(f)}$ for stochastic GW
backgrounds for the
NANOGrav 11-year pulsars is given in Figure~\ref{f:ng11yr_gwb}.
The three curves show the effect of including a contribution
from the GWB to the auto-power spectra of all the pulsars
(blue versus dashed-orange curves) and the false improvement in
sensitivity that arises if one fails to include the red-noise
component of the individual pulsar noise covariance matrices
(green versus dashed-orange curves).
Typical PTA sensitivity curves that one sees in the literature
incorrectly ignore this red noise component.

\subsubsection{Comparing stochastic and deterministic sensitivity curves}
\label{s:compare_det_stoch}

\begin{figure}[h!tbp]
\centering
\includegraphics[width=\columnwidth]{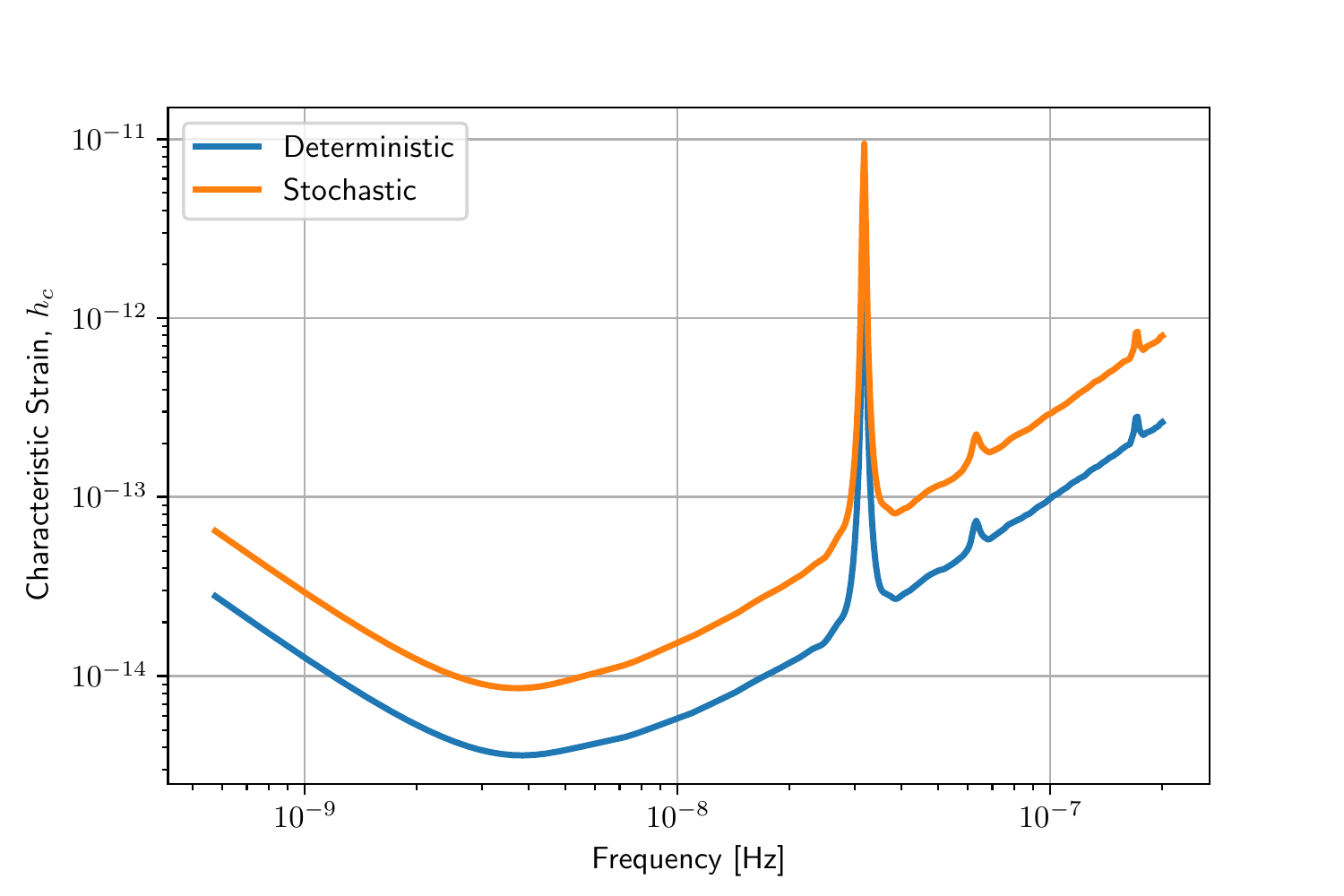}
\caption{Comparison of the sensitivity curves for the 
NANOGrav 11-year pulsars to a single deterministic GW 
signal and a stochastic GW background; see also
Figures~\ref{f:Seff_inc_sky} and \ref{f:ng11yr_gwb}.
The two curves differ by a factor of $\sim2.6$.} 
\label{f:compare_det_stoch}
\end{figure}

Although one uses different statistics to search for
deterministic and stochastic GW signals, it is
interesting to compare the sensitivity curves for
these two different cases.
Figure~\ref{f:compare_det_stoch} shows plots of the
deterministic and stochastic sensitivity curves for
the NANOGrav 11-year pulsars
(taken from Figure~\ref{f:Seff_inc_sky} and
Figure~\ref{f:ng11yr_gwb}, dashed-orange curve).
Note that the sensitivity curve for a single deterministic
source is lower than that for a stochastic background,
since the Hellings and Downs factors $\chi_{IJ}$
in \eqref{e:Seff_gwb}
reduce the effective number of pulsar pairs that
contribute to the stochastic analysis.
To demonstrate this explicitly, compare equations
\eqref{e:Seff_inc,sky} and \eqref{e:Seff_gwb} for $S_{\rm eff}(f)$
assuming that all the pulsars
have the same noise characteristics and timing model fits
(i.e., $S_I(f)\equiv S(f)$ for all $I$), and that all the pulsars
are observed for the full observation time
(i.e., $T_I\equiv T_{IJ}\equiv T_{\rm obs})$.
Then
\begin{align}
&S_{\rm eff}^{\rm det}(f) =\frac{5}{4N_{\rm p}}\,S(f)\,,
\\
&S_{\rm eff}^{\rm stoch}(f) =
\left(\sum_I\sum_{J>I}\chi_{IJ}^2\right)^{-1/2}S(f)\,,
\end{align}
where $N_{\rm p}$ is the number of pulsars.
Since the maximum value of $\chi_{IJ}$ for any pair of
pulsars is $1/2$, we have
\be
\sum_I\sum_{J>I}\chi_{IJ}^2 \le
\frac{N_{\rm p}(N_{\rm p}-1)}{2}\frac{1}{4}\,,
\quad
\ee
which implies
\be
\left(\sum_I\sum_{J>I}\chi_{IJ}^2\right)^{-1/2}
>\frac{2\sqrt{2}}{N_{\rm p}}\,,
\ee
Thus,
\be
S_{\rm eff}^{\rm stoch}(f)
>\frac{2\sqrt{2}}{N_{\rm p}}\,S(f)
\ \Rightarrow\ 
S_{\rm eff}^{\rm stoch}(f)>
S_{\rm eff}^{\rm det}(f)\,.
\ee
Although we have compared the full sensitivity curves
$S_{\rm eff}(f)$
for deterministic and stochatic GW sources,
we note that the corresponding signal-to-noise ratio
for a monochromatic deterministic source uses only the
value of the sensitivity curve at a single frequency $f=f_0$
(see \eqref{e:bar_rho});
while that for a stochastic source involves an integral
of $S_{\rm eff}(f)$ over all $f$
(see \eqref{e:rho2_gwb}
and the discussion in Section~\ref{s:PIcurve}).

\subsubsection{Pairwise stochastic sensitivity curves}
\label{s:pairwise_gwb}

\begin{figure}[h!tbp]
\centering
\includegraphics[width=\columnwidth]{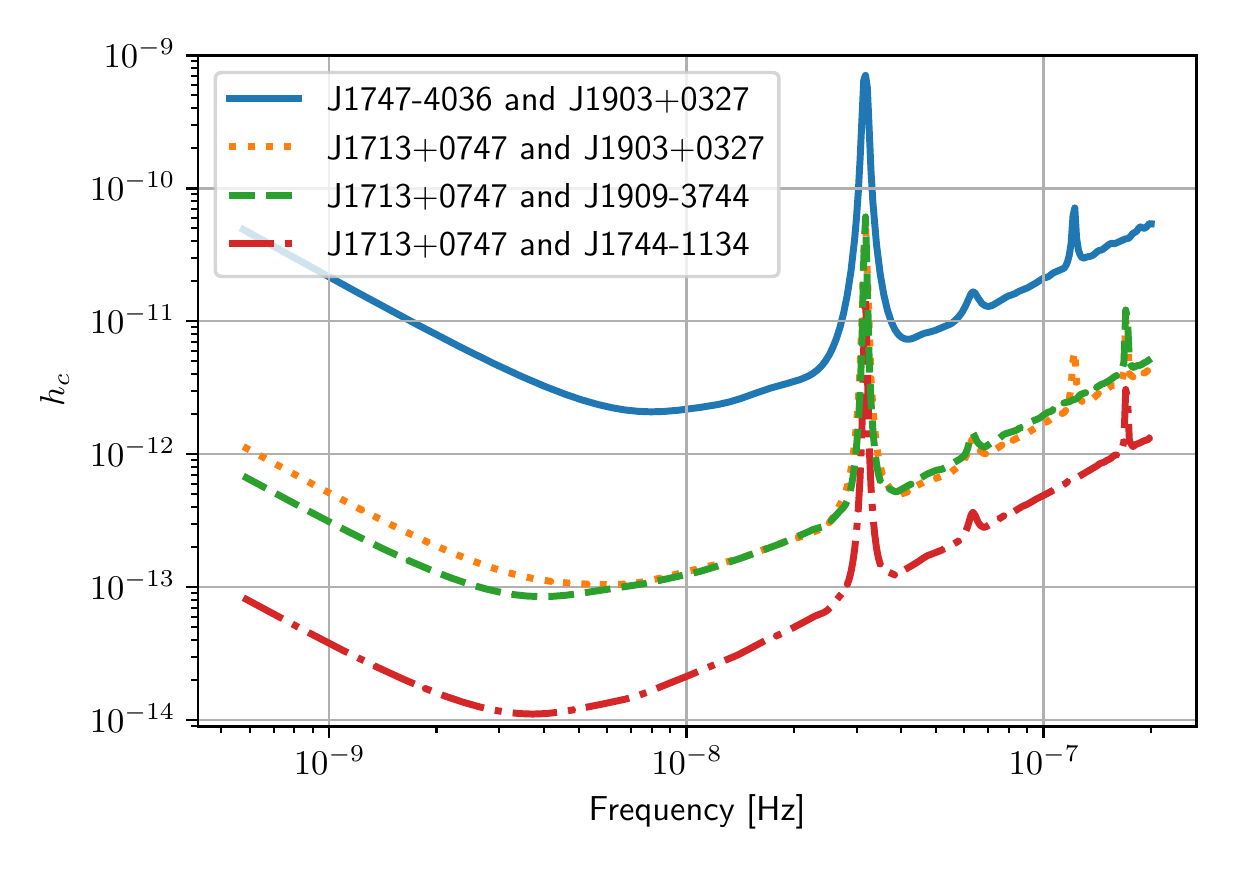}
\caption{Pairwise stochastic sensitivity curves (effective characteristic
strain noise) for a subset of NANOGrav 11-year pulsar pairs.
Since pulsars J1747-4036 and J1903+0327 are two of the
least-sensitive pulsars in the NANOGrav 11-year data set,
their pairwise sensitivity curve is worse (that is, higher)
than the other pairs shown here.
The sensitivity curve for J1713+0747 and J1903+0327 is 
significantly better, since J1713+0747 is the most sensitive
pulsar in the data set; while that for J1713+0747 and J1744-1134
is the best, since both of these pulsars are individually very
sensitive and their Hellings-Downs coefficient is $\chi_{IJ}=0.3304$.
Pulsars J1713+0747 and J1909-3744 are also both individually
very sensitive, but since their Hellings-Downs coefficient is
only $\chi_{IJ}=0.0058$, their pairwise sensitivity curve is
an order of magnitude worse than that for J1713+0747 and J1744-1134.}
\label{f:ng11yr_pulsar_pairs}
\end{figure}

As a by-product of the stochastic sensitivity curve
analysis, we obtain
{\em pairwise} stochastic sensitivity cuves
\be
h_{c,IJ}\equiv\sqrt{f S_{IJ}(f)}\,,
\quad
S_{IJ}(f)\equiv
\sqrt{\frac{T_{\rm obs}}{T_{IJ}}}
\frac{\sqrt{S_I(f)S_J(f)}}{|\chi_{IJ}|}\,,
\ee
by simply restricting ourselves to a single term
in the sum \eqref{e:Seff_gwb}.
Plots of such curves are useful as a diagnostic
for comparing the contribution of different
pulsar pairs to the stochastic optimal statistic
signal-to-noise ratio.
Figure~\ref{f:ng11yr_pulsar_pairs} shows pairwise
sensitivity curves for a subset of the NANOGrav
11-year pulsars, comparing pairwise correlations
of some of the most and least sensitive NANOGrav
pulsars.
%

\subsubsection{Power-law integrated sensitivity curves}
\label{s:PIcurve}

\begin{figure}[h!tbp]
\centering
\includegraphics[width=\columnwidth]{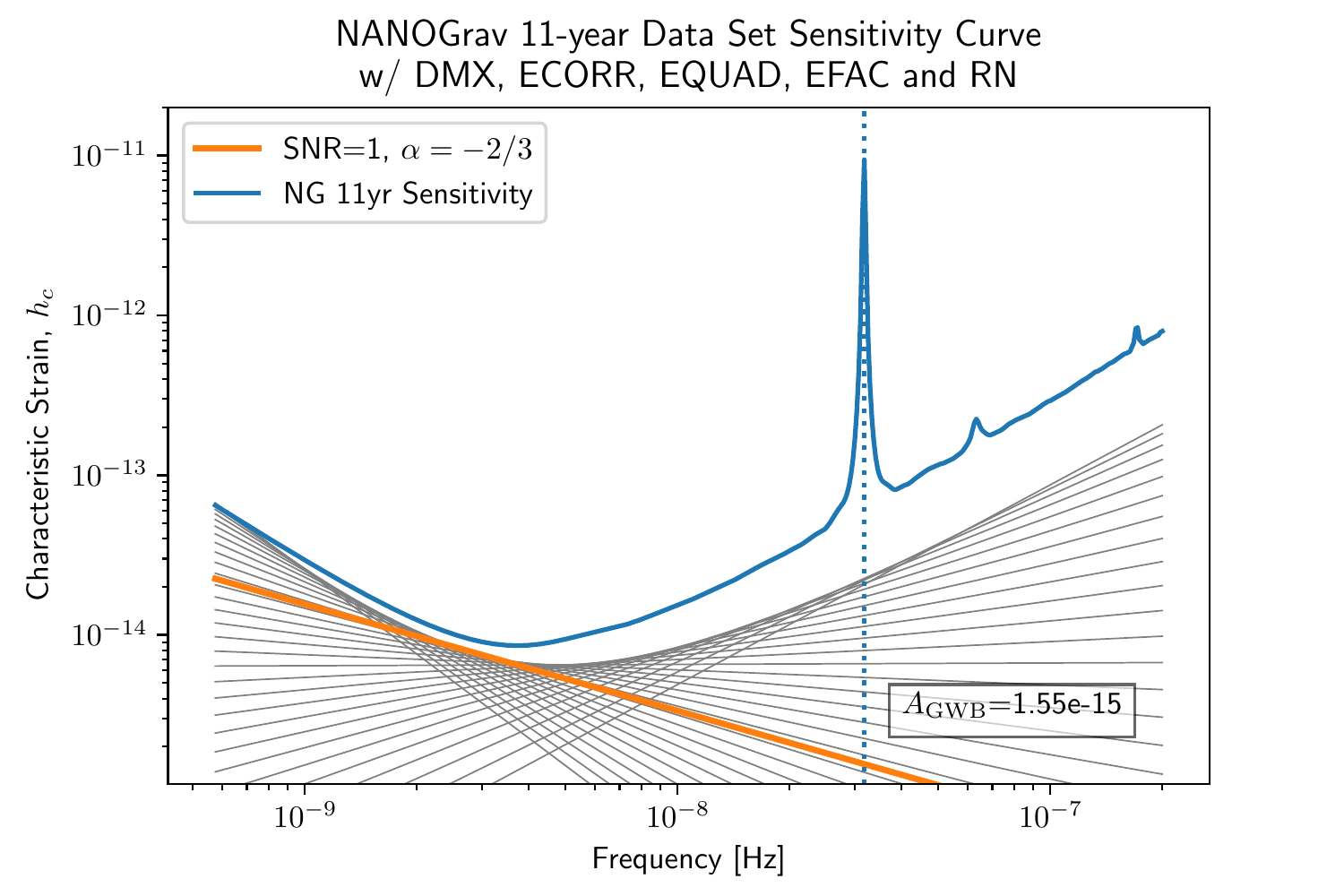}
\caption{Power-law-integrated sensitivity curve for the NANOGrav 11-year
data set. Each of the straight grey lines represents a power-law GWB
detectable with an optimal-statistic signal-to-noise ratio $\rho=1$
for the plotted spectral index.
The envelope of these lines (i.e., the maximum value of all the
power-law backgrounds at a given frequency) defines the
power-law-integrated sensitivity curve for the PTA.}
\label{f:ng11yr_gwb_PI}
\end{figure}

For stochastic backgrounds that have a power-law spectrum,
cf.~\eqref{e:hc_power_law}, it is possible to construct
a sensitivity curve that takes into account the improvement
in sensitivity that comes from integrating over frequency~\cite{Thrane:2013oya}.
Given a range of power-law indices, one
determines the amplitude of each power-law background that
yields a prescribed value of the optimal statistic
signal-to-noise ratio $\rho$ (e.g., $\rho =1$).
The envelope of these power-law backgrounds defines the
power-law-integrated sensitivity curve for the PTA.
Figure~\ref{f:ng11yr_gwb_PI} shows the $\rho=1$
power-law integrated sensitivity curve for the NANOGrav 11-year
data set using the dashed-orange characteristic strain-noise
curve from Figure~\ref{f:ng11yr_gwb}.
%

%% file: manuscript/discussion.tex
\section{Discussion}
\label{s:discussion}

We have presented a method for constructing realistic 
detection sensitivity curves for pulsar timing arrays,
valid for both deterministic and stochastic GW signals.
We can include different noise characteristics and the 
effect of fitting to a timing model via an 
inverse-noise-weighted transmission function 
${\cal N}_I^{-1}(f)\approx {\cal T}_I(f)/{P(f)}$.
Single-pulsar sensitivity curves are then calculated
from the strain-noise power spectral density
$S_I(f) \equiv 1/({\cal N}^{-1}_I(f) {\cal R}(f))$,
where ${\cal R}(f)$ is the polarization and sky-averaged
timing residual response of a pulsar to a passing GW.
Detection sensitivity curves for multiple pulsars (i.e.,
a PTA) are similary constructed from an effective 
strain-noise power spectral density $S_{\rm eff}(f)$, which is 
a combination of single-pulsar strain-noise power 
spectral densities $S_I(f)$, cf.~\eqref{e:Seff_inc,sky}, 
\eqref{e:Seff_inc}, \eqref{e:Seff_gwb}, appropriate
for the GW source that one is interested in detecting.

The sensitivity curves that we have calculated can be 
used to assess the detectability of different GW 
signals by exisiting or planned PTAs. 
The computational cost of producing these sensitivity 
curves is minimal; they can be calculated much faster 
than doing Monte Carlo simulations using injected signals.
By properly incorporating realistic noise 
properties and the effect of timing model fits into
the sensitivity curves, we can produce detectability
estimates that agree quite well with the 
more-computationally-involved predictions.

%% file: manuscript/acknowledgements.tex
\section*{Acknowledgements}
\label{s:acknowledgements}

JSH and JDR acknowledge subawards from the University of Wisconsin-Milwaukee
for the NSF NANOGrav Physics Frontier Center (NSF PFC-1430284).
JDR also acknowledges support from start-up funds from Texas Tech University. TLS acknowledges support from NASA 80NSSC18K0728 and the Hungerford Fund at Swarthmore College.  
Finally, we thank Robert Caldwell, Rutger van~Haasteren and Xavi Siemens for useful discussions and Justin Ellis for sharing some preliminary code.

%% file: manuscript/appendix.tex
\section{Casting the Blandford et al.~analysis~\cite{Blandford:1984a} 
in more modern notation}
\label{s:appendix}

When using pulsar timing data to search for GWs, one needs
to take into account the effects of fitting to a deterministic timing model 
when doing any type of additional signal analysis. 
Following \cite{Blandford:1984a}, we define the residuals
$R\left(t\right)$ as the difference between the observed 
arrival times of the pulses and
the expected arrival times as determined by our best guesses to the 
parameters.
These residuals are fit to an expression linear in the corrections to the 
unknown parameters, $\alpha_{a}$.%
\footnote{In our notation, $R(t)$ is $\delta t_i$ and 
$\alpha_a$ is $\delta\xi_a$.}
(Noise terms are added later in their analysis.)
We start in the notation of \cite{Blandford:1984a}, and then translate to 
expressions in terms of modern PTA GW analyses:
\be
R\left(t\right) =\sum_{a=1}^{N_{\rm par}}\alpha_{a}\psi_{a}\left(t\right)\,.
\ee
We will define $R_i\equiv R\left(t_{i}\right)$, which is a vector
of length $N$, 
and $\psi_{ia}\equiv \psi_{a}\left(t_i\right)$, which is a 
2-dimensional matrix with dimensions $N\times N_{{\rm par}}$. 
(Note we have reversed the order of the indices on $\psi_{ia}$ from
that in \cite{Blandford:1984a}, to be consistent with later work.) 
In more modern PTA data analysis papers, like \cite{vanHaasteren:2012hj} or
\cite{Ellis:2013nrb,Chamberlin:2014ria}, this matrix is referred to as 
the design matrix of the timing model (our $M_{ia}$.) 
The above expression for the residuals can be transformed into an orthonormal 
basis 
\be
R_{i} =\sum_{a=1}^{N_{\rm par}}\alpha_{a}^{\prime}\psi_{ia}^{\prime},
\qquad\psi_{ia}^{\prime}\equiv\sum_{b=1}^{N_{\rm par}}\psi_{ib}L_{ba}\,,
\ee
where 
\be
\sum_{i=1}^{N}\psi_{ai}^{\prime T}\psi_{ib}^{\prime} =\delta_{ab}\,.
\ee
Using these definitions we calculate a relation that will
be useful in the next section. 
To simplify the notation a bit we will use the Einstein 
convention of summing over repeated indices 
without including summation symbols,
using matrix transposes where necessary. 
Thus, for example, the orthonormality conditions can 
be written as
\be
\delta_{ab} =\psi_{ai}^{\prime T}\psi_{ib}^{\prime}
 =L_{ac}^{T}\psi_{ci}^{T}\psi_{id}L_{db}\,.
\ee
Since a change of basis change is invertible, we can act with 
the inverse transformation matrices:
\be
\begin{aligned}
L^{-T}_{ea}\delta_{ab}L^{-1}_{bf} 
&=L_{ea}^{-T}L_{ac}^{T}\psi_{ci}^{T}\psi_{id}L_{db}L_{bf}^{-1}\,,
\\
L^{-T}_{ea} L_{af}^{-1} 
&=\psi_{ei}^{T}\psi_{if}\,,
\end{aligned}
\ee
where $L^{-T}$ denotes the inverse of the transpose matrix $L^T$,
which is the same as the transpose of the inverse matrix $L^{-1}$.
Finally, using the well-known identity for the inverse of a 
product of two matrices:
\be
\left(L_{ea}^{-T}L_{af}^{-1}\right)^{-1} =\left(\psi_{ei}^{T}\psi_{if}\right)^{-1}
\ \Rightarrow\ 
L_{fa}L_{ae}^{T} =\left(\psi_{ei}^{T}\psi_{if}\right)^{-1}\,.
\ee

\subsection{Least-squares regression}
\label{s:least-squares}

One finds the best fit to a timing model by minimizing a 
$\chi^{2}$ function, which we will define below.
In \cite{Blandford:1984a} an ordinary least squares (OLS) 
minimization is used.
In subsequent PTA papers a weighted-least-squares (WLS) 
regression is used, where each residual is weighted by the 
inverse of the TOA error, $W_i\equiv 1/\sigma_i$.
In the most modern work a generalized least squares (GLS) regression is
used where the noise covariance matrix, $N_{ij}$, is used, 
encoding covariances between all residuals:
\be
\begin{aligned}
\chi^{2} \equiv
\left(R_{i}-\alpha_a^{\prime}\psi_{ai}^{\prime T}\right)
N_{ij}^{-1}\left(R_{j}-\psi_{jb}^{\prime}\alpha_{b}^{\prime}\right)\,.
\end{aligned}
\ee
Here we solve the GLS minimization problem, restricting
to simpler scenarios if needed---i.e.,
$N^{-1}_{ij}=\sigma_i^{-2}\,\delta_{ij}$ for the case of WLS, 
and $N^{-1}_{ij}=\delta_{ij}$ for OLS (as noise is not taken 
into account during the OLS fit).
We minimize the expression
for $\chi^{2}$ above by finding the root(s) of the derivative with
respect to the parameters:
\be
\begin{aligned}
0 &= 
\frac{\partial\chi^{2}}{\partial\alpha_{a}^{\prime}}
\\
&=-\psi_{ai}^{\prime T}N_{ij}^{-1}R_{j}+\psi_{ai}^{\prime T}N_{ij}^{-1}\psi_{jb}^{\prime}\alpha_{b}^{\prime}
+ ({\rm transpose})\,.
\end{aligned}
\ee
Solving for $\alpha_{b}^{\prime}$ gives
\be
\alpha_{b}^{\prime} 
=\left(\psi_{ai}^{\prime T}N_{ij}^{-1}\psi_{jb}^{\prime}\right)^{-1}
\psi_{ak}^{\prime T}N_{kl}^{-1}R_{l}.
\label{e:GLS_fit}
\ee
In \cite{Blandford:1984a}, they consider OLS fitting. 
There the noise is taken into account after the fit, but its existence
is implicit throughout. For instance the difference between the LHS and RHS
side of their Equation~(2.9) would be zero if there was no noise. 
Setting $N_{ij}=\delta_{ij}$ gives
\be
\alpha_{b}^{\prime}
=\left(\psi_{ai}^{\prime T}\psi_{ib}^{\prime}\right)^{-1}\psi_{aj}^{\prime T}R_{j}
=\delta_{ab}^{-1}\,\psi_{aj}^{\prime T}R_{j}\\
=\psi_{bi}^{\prime T}R_{i}\,.
\label{e:OLS_fit}
\ee
This is the result that \cite{Blandford:1984a} reports for the best fit. 
For WLS fitting, we have 
\be
\alpha_{b}^{\prime} =\left(\psi_{ai}^{\prime T}W_{ij}^{2}\psi_{jb}^{\prime}\right)^{-1}\psi_{ak}^{\prime T}W_{kl}^{2}R_{l}\,,
\label{e:WLS_fit}
\ee
where $W^2_{ij} \equiv \sigma_i^{-2}\,\delta_{ij}$.

\subsection{Transmission function for ordinary least-squares regression}
\label{s:trans_OLS}

The transmission function is defined by \cite{Blandford:1984a} as the 
transfer function relating the power in the pre-fit residuals $R_i$
to that in the post-fit residuals
\be
R^{\rm post}_i \equiv R_i - \psi^{\prime}_{ia}\alpha_a^{\prime}\,,
\ee
where $\alpha_a^{\prime}$ are the best-fit values to the 
parameter deviations, determined by the $\chi^2$ minimization
procedure discussed above.
For the case of OLS fitting, which \cite{Blandford:1984a} consider,
$\alpha_a^{\prime}$ is given by \eqref{e:OLS_fit}, implying
\be
R^{\rm post}_i 
\equiv R_i - \psi^{\prime}_{ia}\psi^{\prime T}_{aj} R_j
=(\delta_{ij} - \psi^{\prime}_{ia}\psi^{\prime T}_{aj}) R_j\,.
\label{e:OLS_post}
\ee
The variance in the post-fit residual is then
\be
\begin{aligned}
\sigma^2_{\rm post}
&\equiv \frac{1}{N}\langle R^{\rm post}{}^T R^{\rm post}\rangle
\\
&= \frac{1}{N}\langle R_j R_k\rangle
\left(\delta_{ji} - \psi^{\prime}_{ja}\psi^{\prime T}_{ai}\right)
\left(\delta_{ik} - \psi^{\prime}_{ib}\psi^{\prime T}_{bk}\right)
\\
&= \frac{1}{N}\langle R_j R_k\rangle
\left(\delta_{jk} - \psi^{\prime}_{ja}\psi^{\prime T}_{ak}
- \psi^{\prime}_{jb}\psi^{\prime T}_{bk}
+\psi^{\prime}_{ja}\psi^{\prime T}_{ai}\psi^{\prime}_{ib}\psi^{\prime T}_{bk}\right)
\\
&= \frac{1}{N}\langle R_j R_k\rangle
\left(\delta_{jk} - \psi^{\prime}_{ja}\psi^{\prime T}_{ak}\right)\,,
\end{aligned}
\ee
where we used orthogonality of the $\psi^{\prime}_{ja}$ to get the last line.
Since the covariance matrix $\langle R_i R_j\rangle$ is related to 
its power spectral density $P(f)$ via
\be
\langle R_i R_j\rangle = 
\int_0^{\infty}{\rm d}f\> P(f)\, e^{i 2\pi f(t_i-t_j)}\,,
\label{e:cov_psd}
\ee
it follows that
\be
\sigma^2_{\rm post}=\int_0^{\infty}{\rm d}f\> {\cal T}(f) P(f)\,,
\label{e:sigma2_post_int}
\ee
where
\be
\begin{aligned}
{\cal T}\left(f\right) 
&\equiv 1-\frac{1}{N}
\psi_{ia}^{\prime}\psi_{aj}^{\prime T}e^{i2\pi f\left(t_{i}-t_{j}\right)}
\\
&=1-\frac{1}{N}
\tilde{\psi}_{a}^{\prime}\left(f\right)\left(\tilde{\psi}_{a}^{\prime}\left(f\right)\right)^{\dagger}
\label{e:OLS_trans}
\end{aligned}
\ee
with $\tilde{\psi}_a^{\prime}$  the Fourier transforms of the basis 
functions:
\be
\tilde{\psi}_{a}^{\prime}\left(f\right)={\psi}_{ia}^{\prime}e^{i2\pi ft_{i}}. 
\ee
Making this substitution and transforming 
$\psi_{ia}^\prime$ back to the original basis, 
we find
\be
\begin{aligned}
{\cal T}\left(f\right) 
&=1-\frac{1}{N}\psi_{ib}L_{ba}L_{ac}^{T}\psi_{cj}^{T}e^{i2\pi f\left(t_{i}-t_{j}\right)}\\
&=1-\frac{1}{N}\psi_{ib}\left(\psi_{ck}^{T}\psi_{kb}\right)^{-1}\psi_{cj}^{T}e^{i2\pi f\left(t_{i}-t_{j}\right)}\\
&=\frac{1}{N}\left(\delta_{ij}-\psi_{ib}\left(\psi_{ck}^{T}\psi_{kb}\right)^{-1}\psi_{cj}^{T}\right)e^{i2\pi f\left(t_{i}-t_{j}\right)}\,,
\end{aligned}
\ee
which is an expression for transmission function
in terms of the original design matrix $\psi_{ia}$.

\subsection{Transmission function for generalized least-squares regression}
\label{s:trans_GLS}

For the case of GLS fitting, the best-fit values for 
the timing parameter deviations are given by \eqref{e:GLS_fit},
for which the post-fit residuals are given by
\be
\begin{aligned}
R_i^{\rm post}
&= R_i - \psi^{\prime}_{ia}\left(\psi_{bj}^{\prime T}N_{jk}^{-1}\psi_{ka}^{\prime}\right)^{-1}\psi_{bm}^{\prime T}N_{mj}^{-1}R_j\\
&= \left(\delta_{ij} - \psi^{\prime}_{ia}\left(\psi_{bk}^{\prime T}N_{kl}^{-1}\psi_{la}^{\prime}\right)^{-1}
\psi_{bm}^{\prime T}N_{mj}^{-1}\right)R_j \, .
\label{e:GLS_post_primed}
\end{aligned}
\ee
We can write this in terms of the original basis as
\begin{widetext}
\be
\begin{aligned}
R_i^{\rm post}
&=\left(\delta_{ij}-\psi_{ic}L_{ca}\left(L_{be}^{T}\psi_{ek}^{T}N_{kl}^{-1}\psi_{ld}L_{da}\right)^{-1}L_{bf}^{T}\psi_{fm}^{T}N_{mj}^{-1}\right)R_j
\\
&=\left(\delta_{ij}-\psi_{ic}L_{ca}L_{ad}^{-1}\left(\psi_{ek}^{T}N_{kl}^{-1}\psi_{ld}\right)^{-1}L_{eb}^{-T}L_{bf}^{T}\psi_{fm}^{T}N_{mj}^{-1}\right)R_j
\\
&=\left(\delta_{ij}-\psi_{id}\left(\psi_{ek}^{T}N_{kl}^{-1}\psi_{ld}\right)^{-1}\psi_{em}^{T}N_{mj}^{-1}\right)R_j\,,
\label{e:GLS_post}
\end{aligned}
\ee
which has exactly the same form as \eqref{e:GLS_post_primed} with $\psi^\prime_{ia}$ replaced 
by $\psi_{ia}$.
The variance of the post-fit residuals is thus
\be
\begin{aligned}
\sigma^2_{\rm post}
&\equiv \frac{1}{N}\langle R^{\rm post}{}^T R^{\rm post}\rangle
\\
&= \frac{1}{N}
\left(\delta_{ij} - \psi_{ia}\left(\psi_{bl}^{T}N_{lm}^{-1}\psi_{ma}\right)^{-1}
\psi_{bn}^{T}N_{nj}^{-1}\right)
\langle R_j R_k\rangle
\left(\delta_{ki} - N^{-1}_{kq}\psi_{qc}\left(\psi_{dr}^{T}N_{rs}^{-1}\psi_{sc}\right)^{-1}
\psi_{di}^T\right)\,.
\end{aligned}
\ee
Since $\langle R_iR_j\rangle\equiv N_{jk}$ for GLS fitting, we get
\be
\begin{aligned}
\sigma^2_{\rm post}
&= \frac{1}{N}
\left(N_{ik} - \psi_{ia}\left(\psi_{bl}^{T}N_{lm}^{-1}\psi_{ma}\right)^{-1}\psi_{bk}^{T}\right)
\left(\delta_{ki} - N^{-1}_{kq}\psi_{qc}\left(\psi_{dr}^{T}N_{rs}^{-1}\psi_{sc}\right)^{-1}\psi_{di}^T\right)
\\
&= \frac{1}{N}
\left(N_{ik}\delta_{ki} 
- \psi_{ia}\left(\psi_{bl}^{T}N_{lm}^{-1}\psi_{ma}\right)^{-1}\psi_{bi}^T
- \psi_{ic}\left(\psi_{dr}^{T}N_{rs}^{-1}\psi_{sc}\right)^{-1}\psi_{di}^T
\right.
\\
&\left.\qquad\qquad\qquad\qquad\qquad
+ \psi_{ia}\left(\psi_{bl}^{T}N_{lm}^{-1}\psi_{ma}\right)^{-1}
\psi_{bk}^T N_{kq}^{-1} \psi_{qc}\left(\psi_{dr}^{T}N_{rs}^{-1}\psi_{sc}\right)^{-1}\psi_{di}^T
\right)
\\
&= \frac{1}{N}
\left(N_{ik}\delta_{ki} 
- \psi_{ia}\left(\psi_{bl}^{T}N_{lm}^{-1}\psi_{ma}\right)^{-1}\psi_{bi}^T\right)
\\
&= \frac{1}{N}
\left(\delta_{ij} - \psi_{ia}\left(\psi_{bl}^{T}N_{lm}^{-1}\psi_{ma}\right)^{-1}\psi_{bk}^T N^{-1}_{kj}\right)
N_{ij}\,,
\end{aligned}
\ee
where we used the symmetry of $N_{ij}$ throughout.
Finally, using \eqref{e:cov_psd} for $N_{ij}$, 
we recover \eqref{e:sigma2_post_int} with
\be
{\cal T}\left(f\right)=
{\cal T}_R\left(f\right)\equiv 
\frac{1}{N}
\left(\delta_{ij} - 
\psi_{ia}\left(\psi_{bl}^{T}N_{lm}^{-1}\psi_{ma}\right)^{-1}\psi_{bk}^T N^{-1}_{kj}\right)\,
e^{i2\pi f\left(t_{i}-t_{j}\right)}\,.
\ee
We thus obtain the same $R$-matrix-dependent transmission function ${\cal T}_R(f)$
found in \eqref{e:transmission_R},
with the $R$-matrix given by the expression in parentheses,
$R_{ij} \equiv
\delta_{ij} - \psi_{ia}\left(\psi_{bl}^{T}N_{lm}^{-1}\psi_{ma}\right)^{-1}
\psi_{bk}^T N^{-1}_{kj}$.

\end{widetext}